\documentclass{article}
\usepackage{amsmath}
\usepackage{amssymb}
\usepackage{changes}
\usepackage{enumitem}
\setlist{itemsep=3pt}
\usepackage{graphicx}
\usepackage{pdflscape}
\usepackage{multirow}

\usepackage[figurename=Fig.]{caption}
\usepackage{hyperref}
\usepackage{amsmath}
   \usepackage[T1]{fontenc}
   \usepackage[utf8]{inputenc}
\usepackage{textcomp}
\usepackage{lmodern}

\DeclareUnicodeCharacter{0301}{\'}
\DeclareUnicodeCharacter{030B}{\H}

 \usepackage[backend=biber,style=numeric,giveninits=true,citestyle=numeric-comp,natbib=true,doi=true,url=false]{biblatex}

\DeclareFieldFormat{doi}{%
  \href{https://#1}{#1}}

\renewbibmacro{in:}{}

 \renewbibmacro{in:}{%
  \ifentrytype{article}
    {}
    {\bibstring{in}%
     \printunit{\intitlepunct}}}

\newcommand\etal{\textit{et al.}}
\newcommand\x{\times}

\graphicspath{{graphics/}}

\title{Leveraging advances in machine
learning for the robust
classification and interpretation of
networks.}
\author{Raima Carol Appaw$^{1\ast}$, Nicholas Fountain-Jones$^{1}$, Michael A.\ Charleston$^{1}$\\$^{1}$ School of Natural Sciences, University of Tasmania, Hobart Australia 7001. \\$^{}$ }

\begin{document}
\maketitle
\begin{abstract}
The ability to simulate realistic networks based on empirical data is an important task across scientific disciplines, from epidemiology to computer science. 
Often simulation approaches involve selecting a suitable network generative model such as Erdös-Rényi or small-world.
However, few tools are available to quantify if a particular generative model is suitable for capturing a given network structure or organization.
We utilize advances in interpretable machine learning to classify simulated networks by our generative models based on various network attributes, using both primary features and their interactions.
Our study underscores the significance of specific network features and their interactions in distinguishing generative models, comprehending complex network structures, and the formation of real-world networks.
\end{abstract}

\section{Introduction} 
\label{sec:introduction}
Real-world network data derived from physical systems such as ecological food webs, biochemical pathways, genetic interactions, animal social behavior, and biological processes, captures complex relationships and addresses fundamental questions about species adaptability, ecosystem dynamics, pathogen dynamics, social dynamics, and genetic regulatory networks \cite{keeling2011networks,larsen2018selecting, bonner2016deep,ames2011a,fountain2019make,fountain2023spectral}.
The multi-dimensional nature and dynamic interactions among variables over time in these systems pose a challenge to their classification.
Traditional classification methods (such as decision trees, support vector machines, k-nearest neighbor, and logistic regression) struggle to capture these complexities effectively \cite{seary2003spectral,vstrumbelj2014explaining,aliakbary2015distance,kajdanowicz2016using}.
Moreover, the lack of interpretability in machine learning models further compounds this challenge. 

Machine learning techniques, including supervised learning, have been applied to the classification of real-world networks into theoretical network models, also known as generative models.
Jansen \etal\ \cite{janssen2012model} introduced the Alternating Decision Tree (ADT), an unsupervised algorithm designed for network classification by focusing on model selection through synthetic network fitting to real network data. 
Their approach involved generating 1000 graphs and extracting various network features, training the ADT using nine classifiers for effective model selection. 
Similarly, Ikehara \etal\ \cite{ikehara2017characterizing} conducted a study on 986 real networks and 575 generated networks, utilizing binary classification with graph features. 
Barnett \etal\ \cite{barnett2019a} employed a random forest algorithm for network classification, manually selecting individual graph features, while Canning \etal \cite{canning2017network} proposed a random forest approach for classifying 529 networks.
Although these efforts to address network similarity, feature identification, and optimization, the studies did not consider the impact of feature interaction on classification. 
Graph similarity metrics, such as degree distribution and clustering, have been utilized to compare graph centrality measures between real-world and theoretical networks \cite{motallebi2013generative,aliakbary2015distance,gera2018identifying,kajdanowicz2016using,prvzulj2007biological}. Although some measures show effectiveness in classifying real-world networks, there remains potential for enhancing the efficiency of these methods.

Various network models such as Erdös-Rényi, scale-free, small-world, stochastic-block model, and spatial networks, have been used to analyze real-world data \cite{erdos1960a,watts1998a,barab1999a,shirley2005a,keeling2011networks,wills2020metrics,newman2003}.
These network models are straightforward to simulate and facilitate the testing of new and diverse network-related hypotheses.
 Using these network models characterized by different network structures and dynamics has set the stage for pioneering research, offering the potential to classify empirical networks effectively \cite{nagy2022a,larsen2018selecting,bonner2016deep,bonner2016efficient,bai2019simgnn,attar2017classification}.
 An important inquiry in our study revolves around the suitability of a spatial network model that reproduces structural characteristics observed in real-world networks.
 Despite their potential for estimating vertex similarity from graph structures making them superior to other models for classification purposes \cite{janssen2012model}, spatial models are often underutilized in predictive modeling frameworks.
In a spatial model, individuals are positioned in space, where edge formation depends on the distance between individuals \cite{janssen2012model,keeling2011networks}. 
Spatial representation within networks enables the quantification of social space by identifying communities or individuals with similar characteristics \cite{janssen2012model}. 
While scientists utilize network models to classify real-world networks, the challenge lies in identifying the most effective network model for the classification task. Various methods have been employed to tackle this issue; however, none have systematically considered the interactions between network features in the classification of empirical networks \cite{motallebi2013generative,aliakbary2015distance,gera2018identifying,kajdanowicz2016using,prvzulj2007biological,nagy2022a}.
Understanding feature interaction in predictive modeling allows for a more comprehensive analysis of how different features interact to influence a models final outcome, leading to more accurate predictions and better model interpretability.

Our research leverages advances in recent interpretable machine learning techniques to identify not only novel network features that influence classification outcomes but also to uncover the interactions among these network features. 
This approach aims to enhance our understanding of the generative models shaping network formation, revealing critical features influencing the observed structures in diverse physical systems

In this research, we aim to investigate and determine network structural features specific to underlying generative models and uncover the interactions between these features, in order to better classify networks and interpret models.

To this end, we employ a wide range of theoretical network models, 
each representing different structural properties observed in real-world networks. 
We simulated networks from five different generative models, spanning a wide range of sizes and parameter combinations.
This approach ensured that the simulated networks exhibited key target properties such as sparsity, small-world characteristics, clustering coefficient (transitivity), mean path length, and power-law degree distribution \cite{shirley2005a,keeling2011networks,keeling2005networks}.
As a result, our study has two main contributions:

\begin{itemize}
    \item We identify new and specific features that can predict the different generative models, including spectral measures -- like the (normalized) Fiedler value and spectral radius, in prediction of Erdös-Rényi and Small-World networks respectively;
    \item  We established thresholds for feature values where interactions among features become important for network prediction. 
\end{itemize}

To achieve these goals, we employed recent model interpretation techniques like Shapley Additive Explanations (SHAP) \cite{marcilio2020explanations,lundberg2017unified} and Friedman and Popescu’s H-statistics \cite{friedman2008predictive}.
We hope that our modern approach to investigating these features and interactions for classification will benefit future researchers.

Overall, our approach aims to enhance our understanding of the generative models shaping network formation, revealing critical features and their interaction  influencing the observed structures in diverse physical systems.
We also applied our model to real-world networks \cite{weber2013badger,rossi2016a,watts1998a}, demonstrating the practical utility of our approach.

\subsection{Notation and description of network features} 
\label{sub:notation_and_description_of_network_features}

We introduce the notation used in the description of network features used for classification and provide descriptions below.
A graph is defined as a set of \emph{nodes} (referred to as individuals or actors in contact networks) and \emph{edges} that represent connections between pairs of nodes. Mathematically, a \emph{graph}  $G = (V, E)$ consists of a finite set $V$ of nodes and a set $E$ of pairs of nodes ($v_i,v_j$), known as edges.

Two nodes $v_{1}$ and $v_{2}$ are called \emph{adjacent} if there is an edge connecting them.

The \emph{degree} of a vertex $v$, denoted $deg(v)$, is the number of edges connecting that vertex to other vertices. 
In our studies, we focus mainly on undirected graphs, where edges have no  direction -- that is, edge ($v_i,v_j$) is equivalent to ($v_j,v_i$) for all $i,j$. 
The graph $G$ can be mathematically represented using various types of matrices such as the \emph{adjacency matrix} and the \emph{Laplacian matrix}. 
The adjacency matrix of a graph $G$ with $n$ nodes is represented by an $n \times n$ symmetric matrix $A(G)$ and is defined as:
\[
	A = [a_{ij}] := \left\{\begin{array}{cl}
		1 & \text{if $v_{i}, v_{j}$ are adjacent;}\\
		0 & \text{otherwise.}
	\end{array}\right.
\]

The Laplacian matrix is the difference between the degree matrix and the adjacency matrix, given as $L = D - A$.
Additionally, we denote the normalized Laplacian matrix as $\mathcal{L} = D^{-\frac{1}{2}} L D^{-\frac{1}{2}}$, where $D$ and $D^{-\frac{1}{2}}$ are defined below.

The degree matrix $D(G)$ is an $n \x n$ diagonal matrix of graph $G$ defined as
\[
	D = [d_{ij}] := \left\{\begin{array}{cl}
		deg(v_{i}) & \text{if $i=j$;}\\
		0 & \text{otherwise.}
	\end{array}\right.
\]
A closely related matrix of $D$ is $D^{\frac{-1}{2}}$:
\[
	D^{\frac{-1}{2}} = [d_{ij}] := \left\{\begin{array}{cl}
		\frac{1}{\sqrt{deg(v_{i})}} & \text{if $i=j$;}\\
		0 & \text{otherwise.}
	\end{array}\right.
\]

An important object in graph analysis is the \emph{spectrum} of each of the various matrices associated with $G$.
This is an ordered list of the \emph{eigenvalues} of the matrix $M$ in question, written $\lambda_{1}\leq \lambda_{2}\leq \lambda_{3}\leq \dots\leq \lambda_{n}$.

The $j^{th}$ eigenvalue $\lambda_{j}$ of any of the above matrices encodes information associated with the $j^{th}$ eigenvector, which we denote as $\psi_{j}$.
The second smallest non-trivial eigenvalue of the Laplacian and normalized Laplacian matrix respectively are called the \emph{Fiedler} and \emph{normalized Fiedler} values respectively.
Similarly, the largest eigenvalue of the adjacency matrix is called the \emph{spectral radius}.
These three eigenvalues are important in capturing structural properties of complex networks and graphs and have been studied in the application of graph theory and network analysis in many fields \cite{seary2003spectral,akoglu2015a,gera2018identifying}.

With these foundational concepts defined, we can  describe the graph features necessary for the graph classification (refer to Supplementary material : "Definition of Graph Features" for additional details).

\section{Data generation and feature engineering} 
\label{sec:methods_and_data}

\subsection{Theoretical networks (generative models)} 

\label{ssub:Synthetic_network_data}

We utilized the \emph{igraph} package within the \texttt{R} software to generate approximately 442,142 networks, varying in size from 50 to 1000 nodes. Among these simulated networks, 309,499 were allocated for training purposes, while 132,643 were reserved for testing, using a split ratio of 70:30.

For each network type,
we generated over 200 instances for each parameter combination.
We achieved this by adjusting the configuration algorithms and/or model parameters for different network sizes, allowing us to simulate networks with various numbers of edges.
Some network models had wider parameter ranges, resulting in class imbalance (some networks have more simulated data than others). 
To address the class imbalance, we applied the Synthetic Minority Over-sampling Technique (SMOTE) with the \emph{tidymodel} framework in \texttt{R} software \cite{ikehara2017characterizing,kuhn2022tidy,chawla2002smote}.
This involves randomly reconstructing new network instances for all minority classes to equalize the data points, aligning them with the largest class \cite{ikehara2017characterizing,kuhn2022tidy}.
After applying SMOTE, the training data totaled 674,885.
Overall, we had 807,528 networks across the five generative models used for  classification purpose.

For a breakdown of parameter ranges for the generative models, please refer to Table S1.
The generative models are described below.

\subsubsection{Erdös-Rényi Random Graph} 
\label{ssub:random_graph}
We synthesized random graphs using the Erdös-Rényi (ER) graph model \cite{erdos1960a}.
In ER graphs, each node has a fixed probability $P^{(ER)}$ of connecting with every other node in the graph \cite{erdos1960a}. 
For our random graphs, we assigned the connection probability $P^{(ER)}$ drawn from a uniform distribution: $P^{(ER)} \sim U(0.1, 0.9)$.
 In ER networks, the spatial position of individuals is irrelevant with connections forming at uniform random \cite{keeling2005networks}. 
 Therefore, these networks tend to be homogeneous and characterized by low clustering coefficient (transitivity) and short path lengths \cite{keeling2005networks}.
 The degree distribution of an  Erdös-Rényi random graph follows a Poisson distribution \cite{keeling2005networks}

 \subsubsection{Small-World} 
\label{ssub:small_world}
We generated small-world (SW) networks using the Watts-Strogatz small-world model \cite{watts1998a} within the \emph{igraph} package. 
This involved rewiring edges of a regular lattice graph, with rewiring probabilities $P^{(SW)}$ set at 0.1 and 0.3. 
The parameter $l^{s}$, representing the neighborhood within the lattice, ranged from 1 to 35.  
In small-world networks, each edge has a probability $P^{(SW)}$ of being rewired to different random nodes. 
Higher values of $P^{(SW)}$ influence the graph's structure while maintaining the total number of edges.
The small world is characterized by the short path length of the Erdös-Rényi networks and the high clustering of the lattice network \cite{keeling2005networks}.
The degree distribution of the small world network is similar to the Erdös-Rényi network \cite{keeling2005networks}.

\subsubsection{Spatial Graph} 
\label{ssub:spatial_graph}
We constructed a spatial model using a threshold distance parameter, denoted as $r$, in the range of  
 $0.1 \leq r \leq 0.9$.
We randomly distributed $N$ points within a unit square, and pairs of points were connected if their Euclidean distance was within the threshold distance $r$. As $r$ increased, the number of connections in this network also increased.
By altering either the spatial arrangement of individuals in a spatial network, a diverse array of networks can be created, spanning from densely clustered lattices to small world configurations to fully interconnected random networks \cite{keeling2005networks}. 
The degree distribution of spatial networks follows a Poisson distribution \cite{keeling2005networks}.
Spatial networks are characterized by a high degree of heterogeneity \cite{keeling2005networks}.

\subsubsection{Scale-Free} 
\label{ssub:scale_free}
We constructed scale-free (SF) networks using the Barabasi-Albert preferential attachment model \cite{barab1999a}, where nodes are added one at a time. 
In this model, each new node $v_{i}$ attaches up to $m$ pre-existing nodes $v_j$ according to the following probability formula:
$P^{(BA)}$ = $P(v_j)$= $deg(v_{j})^{\alpha}$ / $\sum_{i} deg(v_{i})^{\alpha}$. 
The existing nodes to which the new node is attached are selected randomly, with replacement, using these probabilities, potentially resulting in fewer than $m$. 
Our simulations considered values of $m$ $\in$ $\{1,2,3,4,5,....,35\}$ and the preferential attachment parameter $\alpha$ $\in$ $\{1, 2, 3\}$.
In network analysis, individuals with high connectivity, known as super-spreaders, play a crucial role in propagation or diffusion processes like disease transmission \cite{keeling2005networks}. Including superspreaders in networks is essential for capturing the complexities of propagation processes. Scale free networks offer a solution by enabling extreme levels of heterogeneity \cite{keeling2005networks}.
Similar to the random networks, in scale free networks, the spatial position of individuals is ignored in forming connections or links \cite{keeling2005networks}.
Their degree distribution follows a power-law pattern \cite{keeling2005networks,barab1999a}.

\subsubsection{Stochastic-Block-Model} 
\label{ssub:stochastic_block_model}
Stochastic block models (SBM) extend random graph models by introducing explicit community structures \cite{wills2020metrics,newman2003,nagy2022a}. 
In SBMs, nodes are partitioned into communities or `blocks,' where nodes have stronger connections within the same community than between different communities \cite{wills2020metrics}. 
The parameters of SBMs include the number of nodes $N$, the number of disjoint blocks $s$, each with a size $C_{j}$ for $j=1, \dots, s$, into which the nodes are partitioned, and an $s\times s$ symmetric edge probability matrix $P$. 
This matrix defines the within- and between-community connection probabilities for the nodes. 
For our experiments, we generated graphs under this model with $s=2$ blocks, $C_1 = 0.4N$, $C_2=N-C_1$, and varying $P$. 
Specifically, the connection probability within block $i$ was drawn from a uniform probability density function $U(0.5,0.9)$, and the probability $P$ between blocks was drawn from $U(0.1,0.4)$.

\begin{figure}[htbp]

\begin{centering}
\includegraphics[width=\textwidth]{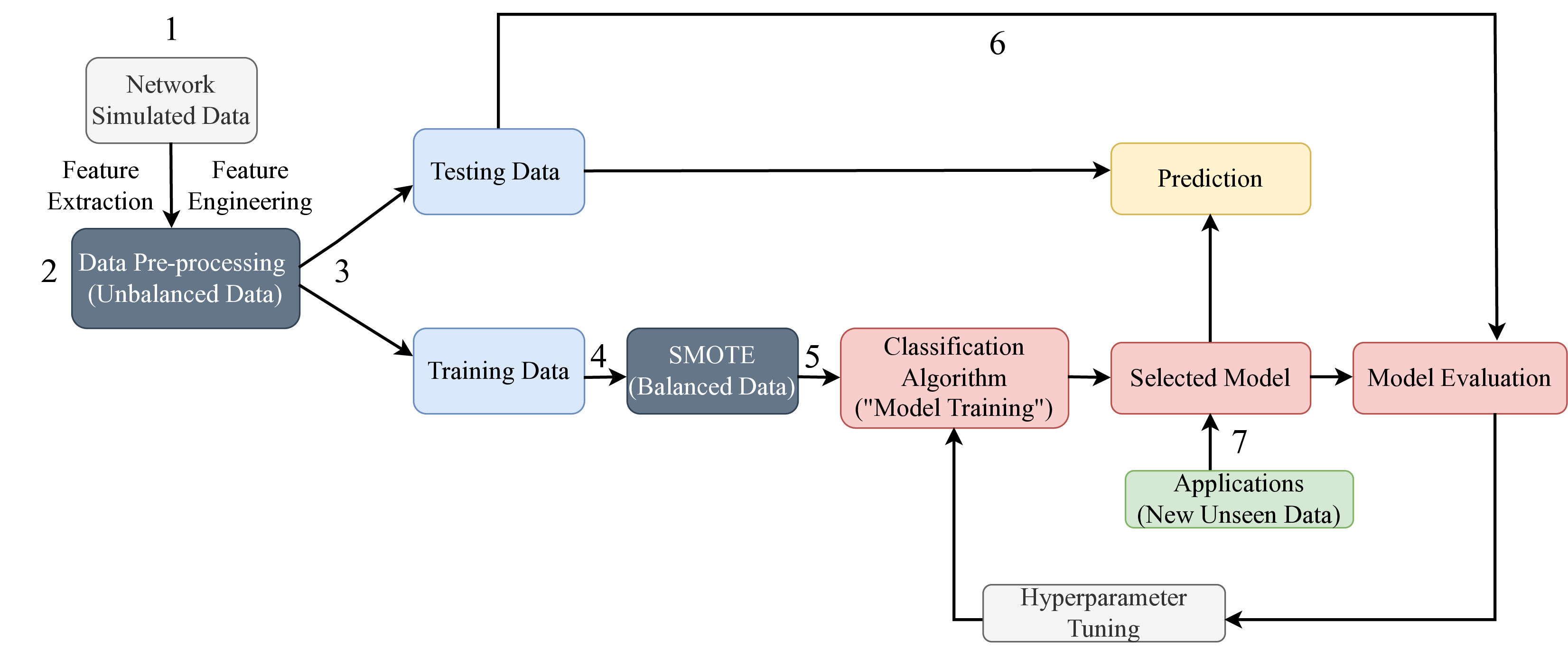}
\end{centering}

\caption{Overview of the network classification method. 
1. Many examples of networks are simulated with the generative models (light grey) and their features are calculated and important features retained during data preprocessing (dark grey). 
2. Data pre-processing includes feature extraction and engineering. 
3. Data is split into test and training data sets (blue); 
4. Minority Over-sampling Technique (SMOTE) with the \emph{tidymodel} framework in \texttt{R} software \cite{kuhn2022tidy} is applied to correct for class imbalance; 
5. Models are trained, and the best classification model is selected (pink), once the models' hyperparameters have been tuned (light grey);
6. The test data is used for model prediction (yellow) and evaluation (pink);
7. The final selected model is applied  to new data sets (green) to predict the generative model class (yellow).}
 \label{fig:flowchart}
\end{figure}

\subsection{Empirical networks (model application)}

The empirical network data used for model application were obtained from published works \cite{weber2013badger,rossi2016a,watts1998a}.
These networks consist of the electrical power grid of the western United States \cite{watts1998a}, the badger social network \cite{weber2013badger}, macromolecular networks such as protein and metabolic
networks \cite{posfai2016network,schellenberger2010bigg}, the biological gene networks, and the distribution of the sub-webs within ecological food web \cite{melian2004food,rossi2016a,almaas2007scale}.

\subsection{Feature-selection}
\label{ssub:features-selection}
We computed graph features for both theoretical and empirical networks and used these features as feature vectors in our classification tasks. 
These features capture various aspects of network structure. 
For a detailed definition of the graph features, refer to Supplementary Material: "Definition of Graph Features". 
In total, we manually selected 18 features, excluding features such as maximum degree, number of triangles, maximum triangle, and total number of triangles, which were correlated with the number of nodes (order).
The 18 selected features are: degree centrality, eigen centrality, closeness centrality, betweenness centrality, degree assortativity coefficient, Fiedler value, normalized Fiedler value, spectral radius, modularity, clustering coefficient (transitivity), mean degree, diameter, mean eccentricity, mean path length, graph energy, minimum cut, order (number of nodes), and number of edges.

\begin{figure}[htbp]
	\begin{center}
    \includegraphics[scale=0.4]{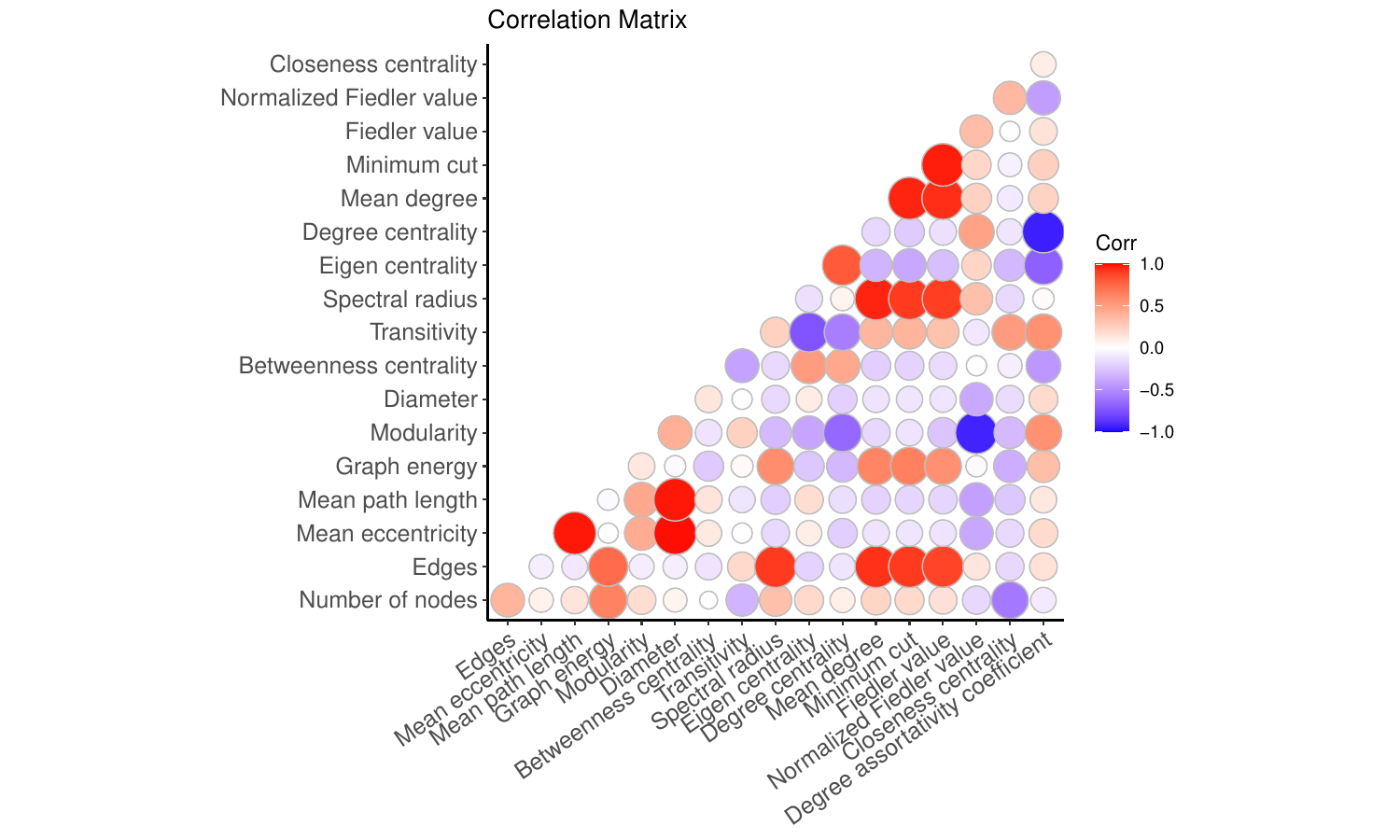}
	\end{center}
	\caption{Correlation among selected network features.
 Most features are correlated (using the Pearson correlation).
 The strength and direction of correlation are indicated by size and color.
 For example, Fiedler value and mean degree are highly positively correlated, and normalized Fiedler has a higher negative correlation with modularity.}
\label{fig:corplot}
\end{figure}

We investigated the interdependence among the selected features using a correlation matrix to detect potential multicollinearity \cite{hollander2013nonparametric,kendall1938new}.
Spearman's Correlation Coefficient ($\rho$) \cite{spearman1961proof,kendall1938new} was utilized to identify the features with high correlations (Fig.~\ref{fig:corplot}).

To address feature correlation and select relevant features for our classification tasks, we employed the `Boruta' feature selection algorithm 
\cite{kursa2010feature}.
Boruta identifies all relevant features by iteratively comparing the importance of each feature with that of a ``shadow'' feature created by permuting the original feature values \cite{kursa2010feature}.
This process helps distinguish truly important features from those that are not \cite{kursa2010feature}.
Despite significant correlations among the features shown in Fig. \ref{fig:corplot}, the Boruta algorithm retained all 18 graph features. 
This suggests that each feature, as well as the collective set, is important for predicting the generative models. Additionally, SHAP values provide further insights by highlighting which features contribute most to the model's final predictions.
This combined approach ensures the retention of all relevant features, avoiding the loss of potentially critical information due to feature correlation, and enhances our ability to interpret the importance of individual features in the model’s final outcome.

\section{Model description and interpretation tools}
\label{sub:experimental_setup}
We employed the \emph{random forest} (RF) and \emph{boosted tree} (XGBM) classification algorithms for training and testing, utilizing features extracted from various network instances and parameters, spanning the generative models. 
 Both RF and XGBM are tree-based algorithms; XGBM fits trees sequentially to correct errors from the previous one, while trees in the RF model are fitted in parallel independently with each iteration \cite{fountain2019make}.
Feature engineering was applied to the extracted features to determine which features to retain or eliminate, enhancing algorithm efficiency, performance, and overall predictive model accuracy \cite{fountain2019make}. 
The training and testing sets were divided into a 70:30 ratio, with both testing and training procedures leveraging the \emph{tidymodel} framework within the \texttt{R} environment \cite{kuhn2022tidy}. 
Tunable hyperparameters included ``mtry”, and ``trees", for RF, and ``mtry", “learning rate”, ``trees", and “tree depth” for XGBM.
 In the \emph{tidymodel} framework, various combinations of hyperparameter values can easily be built using packages such as \emph{tune}, \emph{dials}, and \emph{tune\_race\_anova} \cite{fotache2021high,kuhn2022tidy,kuhn2014futility}. 
 We employed the \emph{grid\_maximum\_entropy} method from the \emph{tune\_race\_anova} package to tune hyperparameters in the \emph{tidymodel} framework, generating a grid search with points selected to maximize entropy for both XGBM and RF models. 
 In total, 400 and 160,000 combinations of hyperparameter values were generated for RF and XGBM, respectively.

In addition, a repeated $k$-fold cross-validation (with $k$ set to 10), randomly splitting datasets into $k$ groups, was utilized in the training process to estimate model performance and evaluate the impact of tuning hyperparameters, such as learning rate, on RF and XGBM model performance \cite{fountain2019make}. This approach aids in selecting the optimal model by considering factors like sensitivity and specificity via the confusion matrix for classification models \cite{fountain2019make}. Cross-validation prevents overfitting and artificial inflation of accuracy by assessing model performance on different data subsets \cite{fountain2019make}.

 \subsection{Shapley Additive Explanations (SHAP) model agnostic interpretation method} 
\label{ssub:SHAP}

SHAP, based on game theory principles, assigns distinct importance weights to individual features according to their contributions to model output and their interactions with other features 
\cite{marcilio2020explanations,lundberg2017unified,vstrumbelj2014explaining}. 
Computation of SHAP values involves solving a system of linear equations using a specialized weighted linear regression approximation method, quantifying how much importance the model attributes to each feature in its predictions \cite{marcilio2020explanations, lundberg2017unified, vstrumbelj2014explaining}.
SHAP analysis provides both local and global feature importance and interpretations, including waterfall plots, variable importance and summary plots, SHAP interaction plots, and dependency plots.
In context, global feature importance assesses each feature's overall contribution to a model's performance or predictions across the entire dataset, while local feature importance evaluates an individual feature's contribution to a single data instance.
SHAP model output is typically expressed in log-odds units \cite{shoham2022predicting}. 
Overall, SHAP values can provide important insights into a models performance in highlighting the most impactful features for each predicted class category \cite{marcilio2020explanations}.

The local and global SHAP techniques employed in our study to investigate the main feature effects are described below:
\begin{enumerate}

\item 
The SHAP waterfall plot serves as a local feature importance visualization, displaying the contribution of each feature to a model's output for a specific data instance. 
This plot helps visualize how individual features influence a particular prediction, aiding in understanding the model's decision-making process. 
The x-axis represents SHAP values' magnitude, indicating the expected predicted values of the model's output. The y-axis, represented by horizontal bars, corresponds to the features and their contributions to shifting the model's output from a reference (base) value. 
Bar length signifies the SHAP value's magnitude for a specific feature, and bar color indicates whether the feature's effect on the prediction pushes it toward a higher (positive, red bars pointing right) or lower (negative, blue bars pointing left) model output.
\item 
The SHAP global feature importance can be presented as a feature summary plot, or global variable importance plot, or a combination of both.
These plots illustrate the direction and magnitude of each feature's impact on a model's final predictions. 
They can be visualized as bars, similar to variable importance bar charts, or as dots representing feature summaries. Dots may be colored to indicate the feature's value (high or low), while bar length corresponds to each feature's strength \cite{zhou2022interpretable, marcilio2020explanations,lundberg2017unified}.
The global feature importance can also be explored with the SHAP dependency plot \cite{zhou2022interpretable, marcilio2020explanations,lundberg2017unified}.

 The SHAP plots identify the most influential model features for individual and global predictions, offering insights into the model's behavior across the dataset \cite{zhou2022interpretable, marcilio2020explanations,lundberg2017unified}. 
 Note that in many cases only a small range of these values is important to classification, such as mean eccentricity for spatial networks and mean path length for the stochastic block model; conversely transivity is important across a wide of values when predicting Erdős-Rényi.
Features can exhibit negative or positive SHAP values, reflecting their impact on the model's prediction. 
High positive SHAP values exert a greater positive impact on the prediction, while large negative SHAP values correspond to negative predictions. 
For example, positive SHAP value for high transitivity and low mean eccentricity tend to favor spatial networks, where as positive SHAP value for low modularity and high normalized Fiedler suggest Erdős-Rényi networks.

\item 
SHAP dependency plots further explore global feature importance, illustrating the relationship between a feature and its impact on model's prediction using scatter plots \cite{zhou2022interpretable, marcilio2020explanations, lundberg2017unified}. These plots complement SHAP feature effect and importance plots by revealing whether a feature has a linear, monotonic, or more complex relationship with the model's prediction. Together, they offer detailed insights into how individual feature values influence the model's final output or prediction. \cite{zhou2022interpretable, marcilio2020explanations, lundberg2017unified}.  

\item 
The SHAP 2D dependency plots are used to explore feature interactions and visualize the combined effects of two features on a model's predictions \cite{lundberg2017unified}. These plots help in identifying and understanding the shape of the joint effects, revealing complex relationships between features. In these plots, point colors can represent different features or the predicted output, providing additional layers of insight \cite{zhou2022interpretable, lundberg2017unified}. This visualization technique is valuable for diagnosing and interpreting the interplay between features, making it easier to see how they jointly influence the model's behavior.
\end{enumerate}

\begin{table}
\caption{Shows performance of our classification pipeline when making predictions with test data. 
These metrics provide insights into how well the model is performing in terms of correctly identifying true positives, true negatives, false positives, and false negatives. 
They are calculated by comparing the model's predictions with the actual ground truth labels in the test dataset.
This figure also shows the proportion of the prediction variability unexplained by the main effect $H^2$ for
    ER: Erdös-Rényi; SBM: Stochastic-block-model; SF: Scale-free; Sp: Spatial; SW: Small-world.}
    \centering
  \begin{tabular}{rrccccccc}
    & \multicolumn{6}{c}{Truth} & \\
      & &  ER & SBM & SF & Sp & SW & Recall &$H^{2} (\%)$ \\
        \hline
    \multirow{5}{*}{\rotatebox{90}{Predicted}} &   ER & 7421 & 5 & 0 & 0 & 13 & 0.9975 & 67\\ 
      & SBM & 0 & 4511 & 0 & 0 & 1 & 0.9997& 80 \\
      & SF & 1 & 0 & 57523 & 0 & 0 & 0.9999& 80 \\ 
      & Sp & 0 & 0 & 0 & 7045 & 0 & 1.0& 80 \\
      & SW & 15 & 9 & 0 & 0 & 56099 & 0.9995& 81\\ 
       \hline
      & Precision:   & 0.9978 & 0.9969 & 1 & 1 & 0.9997 \\
    \end{tabular}
     
     \label{fig:classfic}
\end{table}

\begin{figure} [htpb]
\begin{tabular}{ll}
    (a) \\
    \includegraphics[scale=0.06]{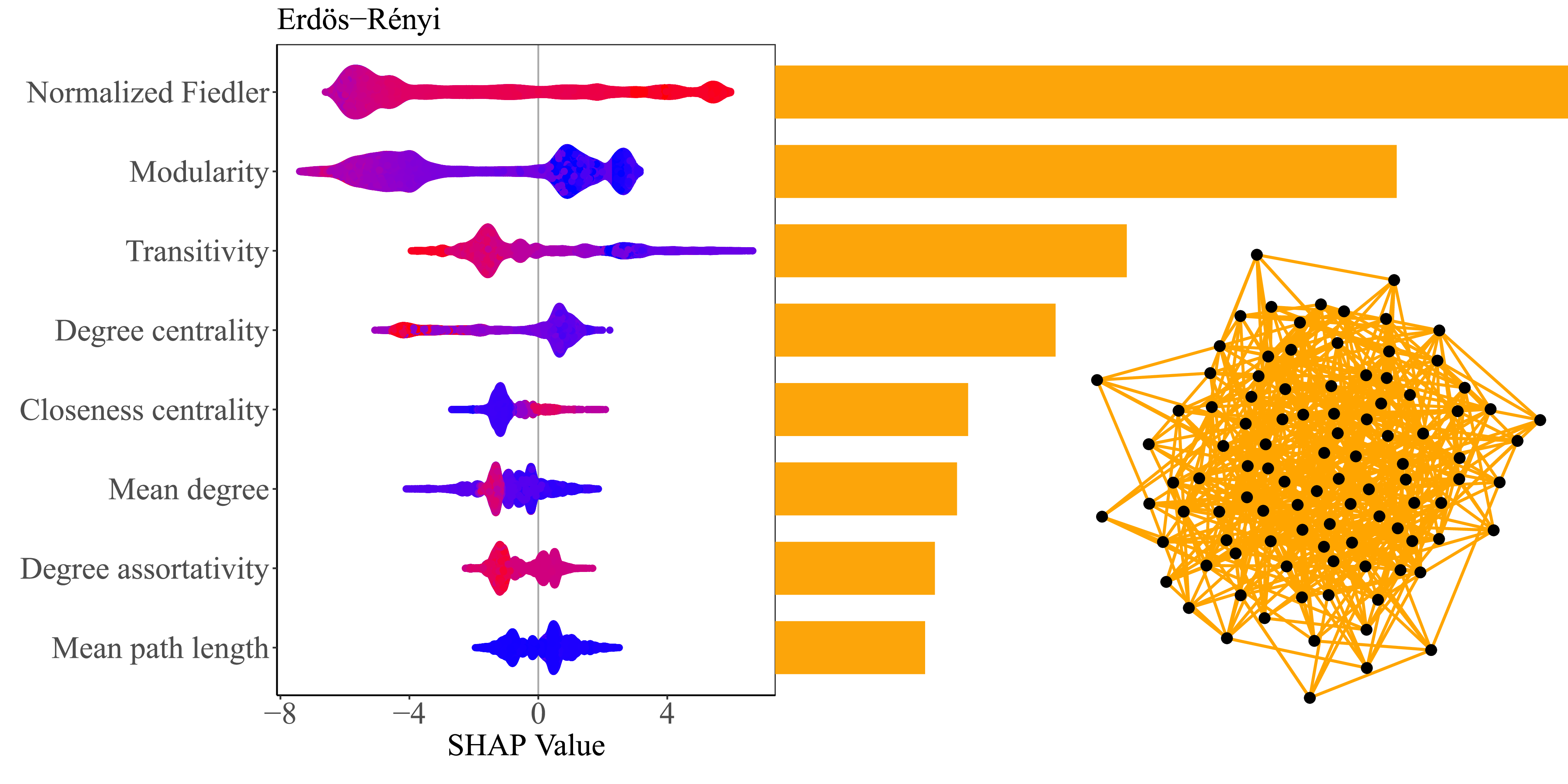} \\
    \includegraphics[scale=0.06]{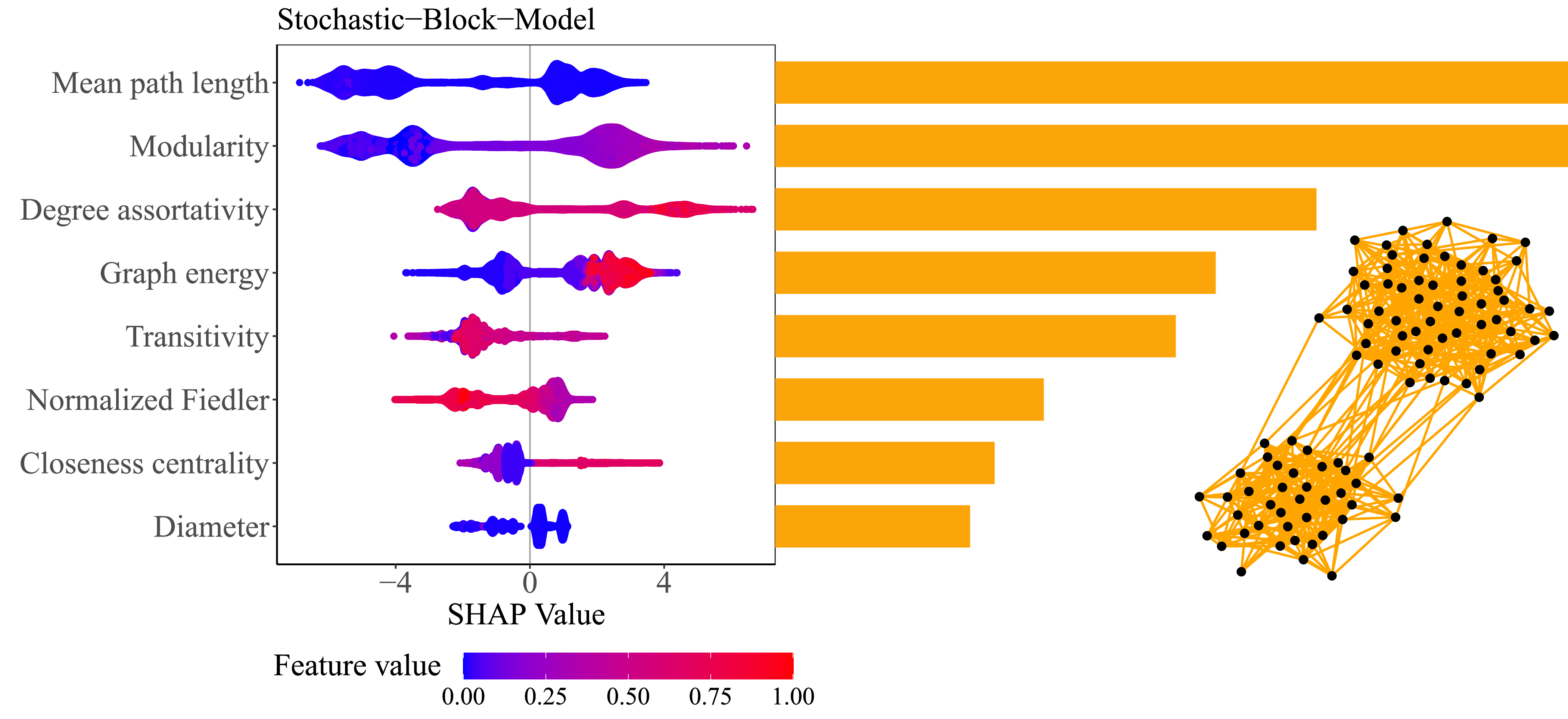}\\
\end{tabular}
\end{figure}

\begin{figure} [htpb]
\begin{tabular}{ll}
    (b) \\
    \includegraphics[scale=0.06]{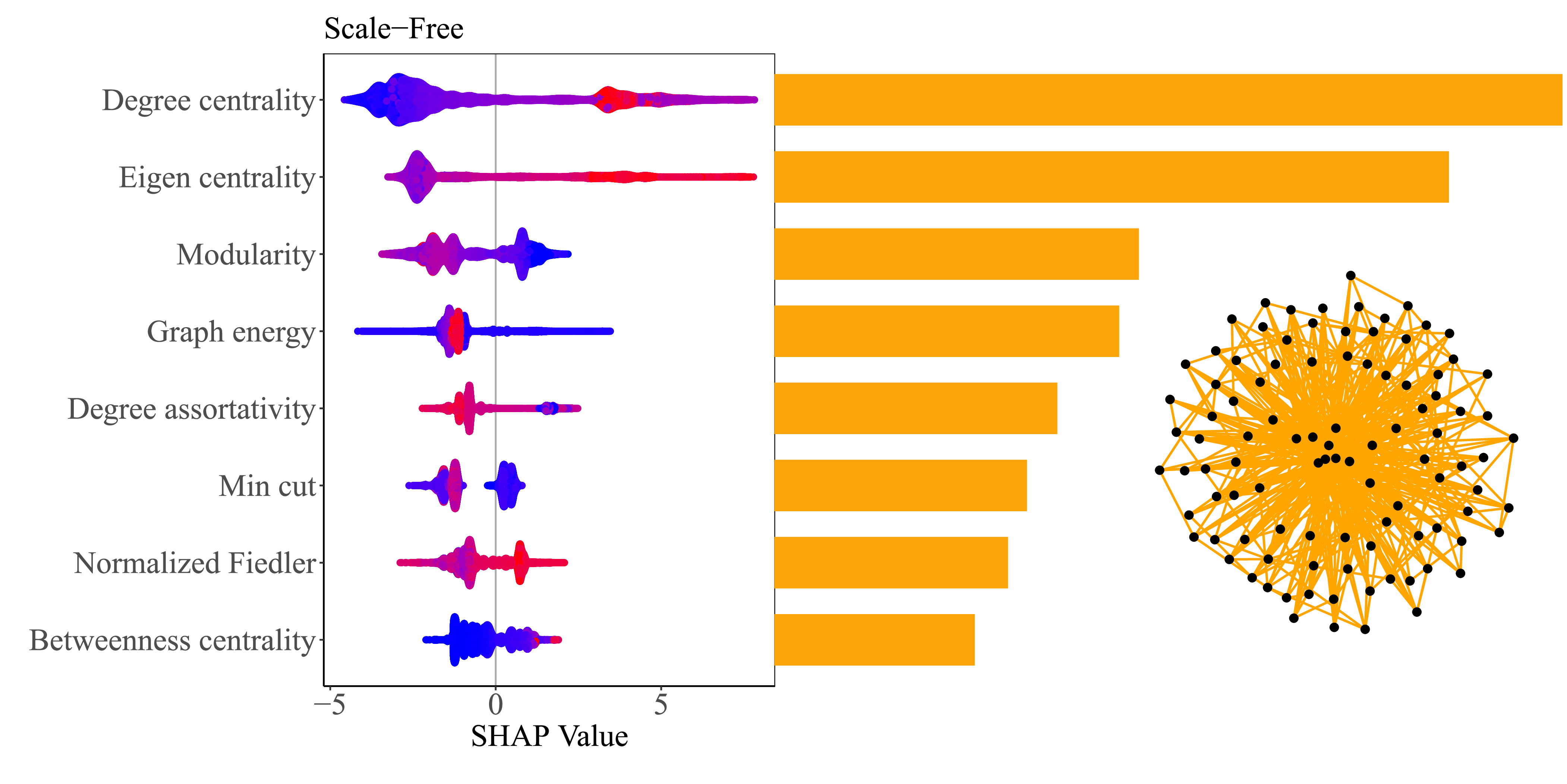}\\
     \includegraphics[scale=0.06]{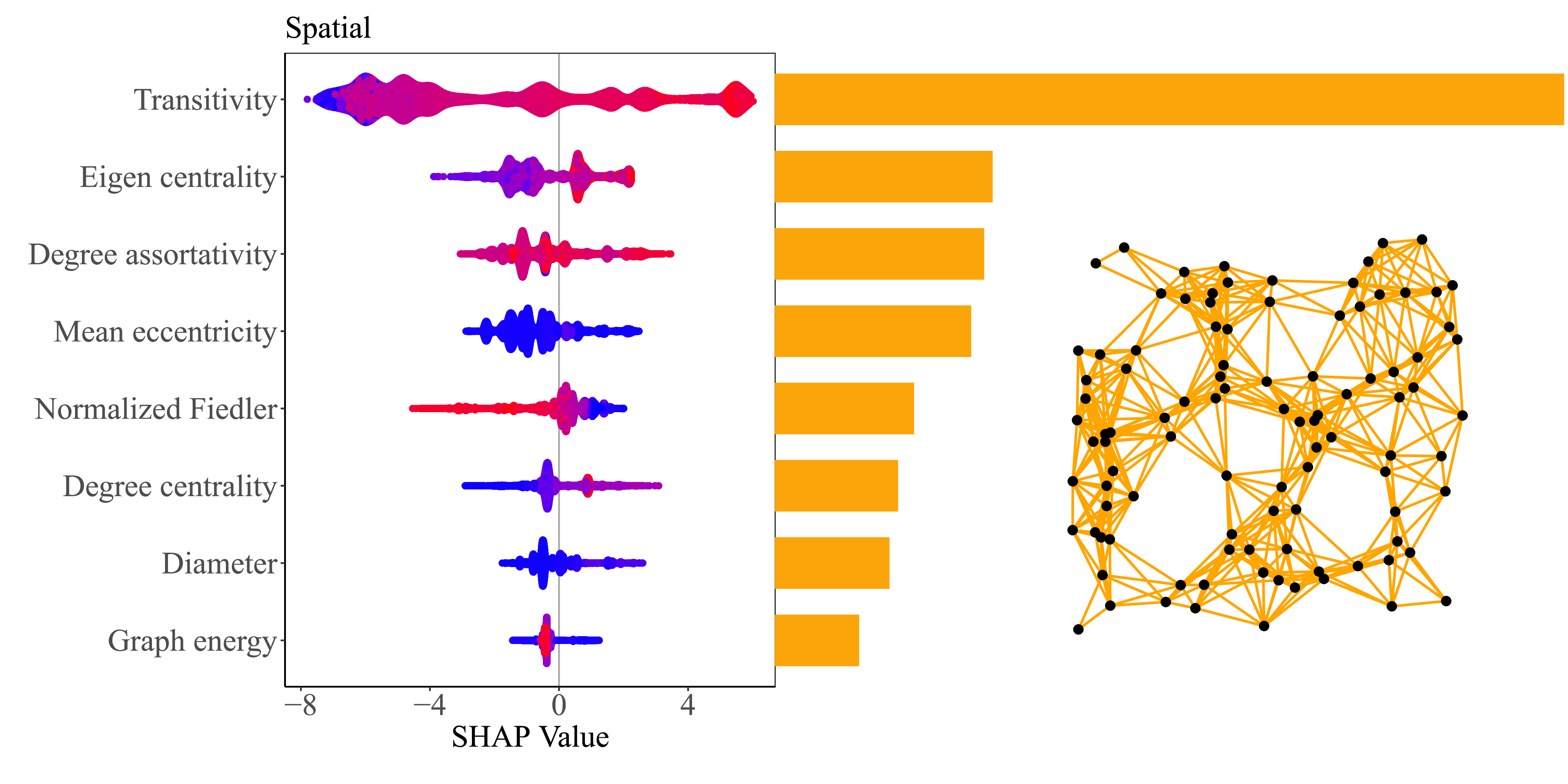} \\
    \includegraphics[scale=0.06]{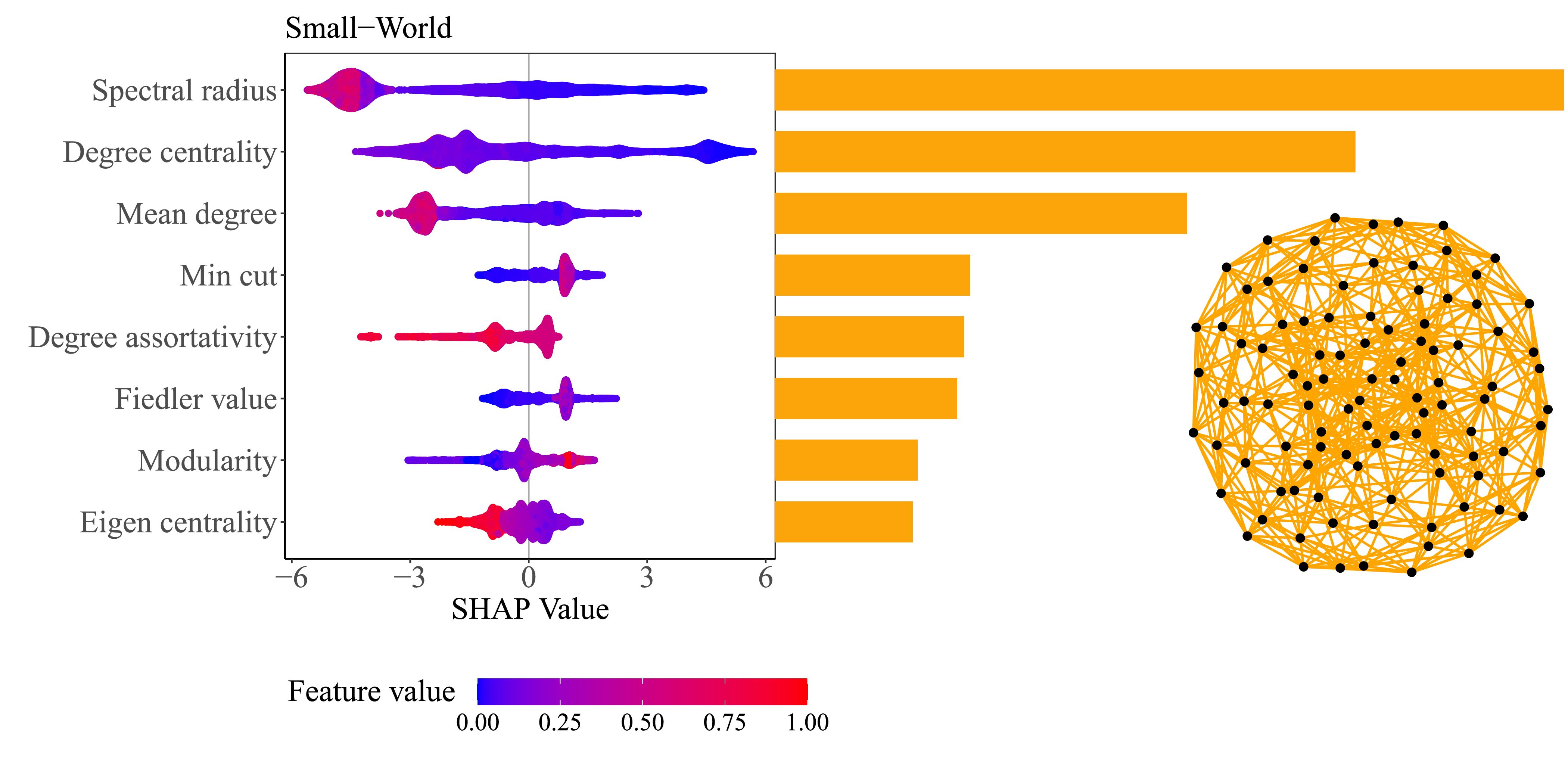} \\
\end{tabular}

\caption{The SHAP feature importance plot provides insight into the significance of various features in predicting Erdös-Rényi, stochastic-block-model, scale-free, spatial, and small-world networks. Key features are positioned at the top of the
plot and are represented by longer orange bars, indicating their stronger importance in influencing the model’s outcome.
Notable predictors for the network types include normalized Fiedler, modularity, degree centrality, transitivity, and spectral
radius. In the feature effect plot, blue indicates lower feature values, while red denotes higher values, across the population
of all network instances.
}
\label{fig:varimp}
\end{figure}

\subsection{H-statistics}
Friedman and Popescu's H-statistics, detailed in \cite{friedman2008predictive}, serve as vital tools for evaluating the performance of machine learning algorithms in classification or regression tasks. 
These statistics identify significant variables or features involved in interactions, quantifying the strength and degree of these interactions, and capturing the key variables influencing a model's predictions.
In addition, we also calculated a global statistic denoted as \textit{H}$^2$, introduced in the same GitHub resource at
\href{https://github.com/mayer79/hstats\#background}
{https://github.com/mayer79/hstats\#background}.
In summary, these measures provide a valuable technique for quantifying and gaining insights into the complex landscape of feature interactions within our classification model.
We described these measures below:
\begin{enumerate}
     \item \textit{H}$^2$:
      Describes the proportion of prediction variability not explained by main effects \cite{zolkowski2023methods}. 
    \item \textit{H}$_i^2$: Describes the proportion of the variability in prediction explained by interactions with feature $i$ (overall interaction effect for individual feature) \cite{friedman2008predictive}.
    \item 
    \textit{H}$_{ij}^2$: Describes the proportion of the variability in prediction explained by the pairwise interaction effect between feature $i$ and $j$ \cite{friedman2008predictive}.
    \item 
    \textit{H}$_{ijk}^2$: Describes the proportion of the variability in prediction explained by the three-way interaction effect among feature $i$, $j$, and $k$ \cite{friedman2008predictive}.
\end{enumerate}

\section{Results} 
\label{sec:results}
Our results demonstrate that both RF and XGBM achieved a nearly 100\% accuracy rate, and an AUC (Area Under the ROC Curve) score of near one across the $k$-fold cross-validation resamples. 
In addition, both specificity and sensitivity were close to 100\%, showcasing the high classification performance of our model.
Further analysis focused on the XGBM due to its superior speed, scalability, and robustness when handling large datasets and real-time prediction tasks compared to RF.

\begin{figure}[htbp]
	\begin{centering}
	\includegraphics[width=0.8\textwidth]{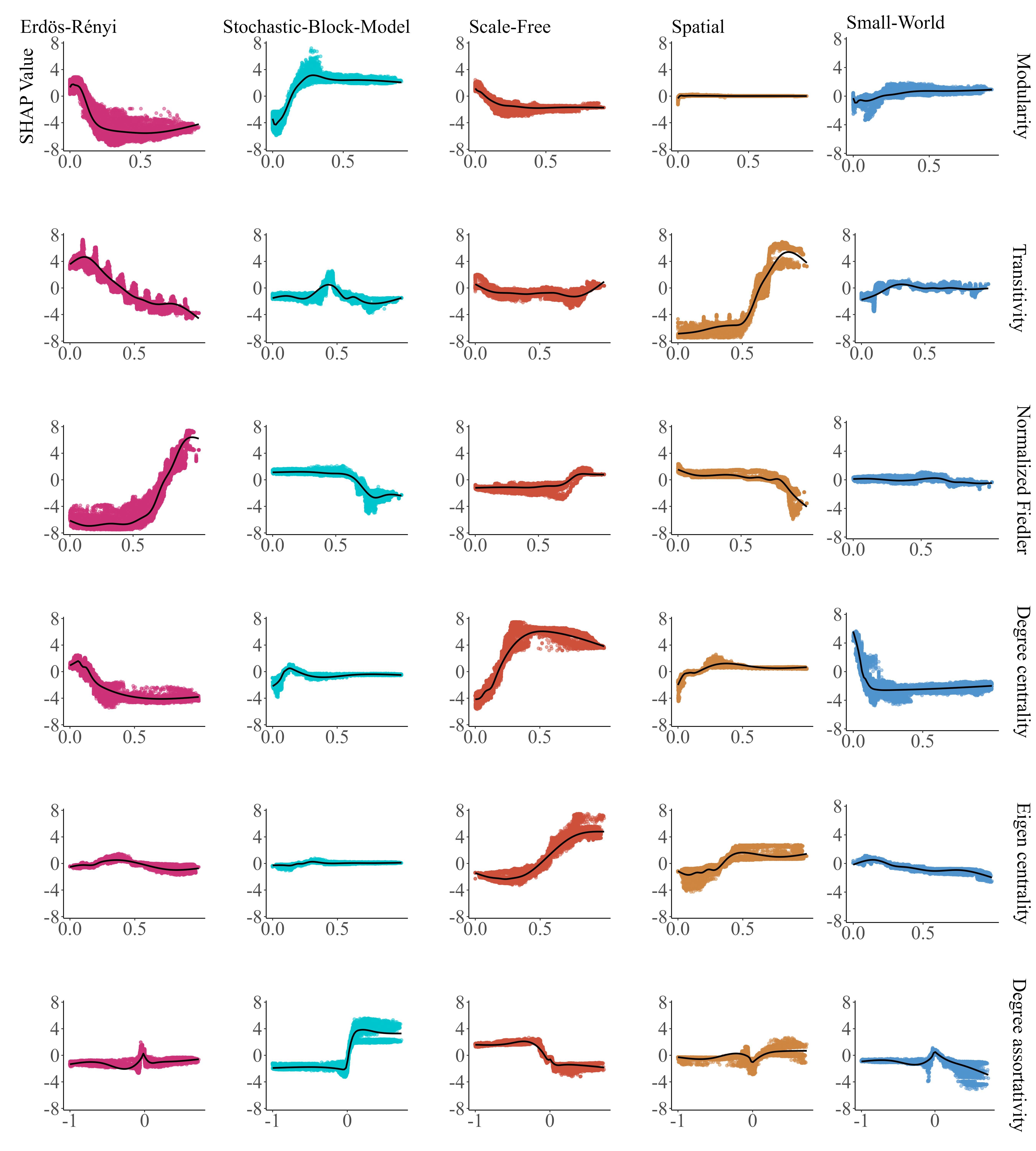}
	\end{centering}
	\caption{SHAP dependency plot showing the scatter plot of the relationship between each feature and its importance in predicting the final output of the Erdös-Rényi, stochastic-block-model, scale-free,  spatial, and small-world generative models across different levels or values. 
 The y-axis typically represents the SHAP value, which quantifies the impact of the feature on the model's prediction, while the x-axis shows the feature's value.
 This plot provides insights into how the model's prediction changes as the feature's value varies, for each class separately.
 The direction and magnitude of the SHAP values across the different classes aid in discerning how important the features are for each class prediction and whether its effect is consistent across all classes or varies. 
 Overall, this plot aids in understanding the model's behavior and the relative importance of features across different classes, providing valuable insights into the model's overall decision-making process for all features across all model classes.}
    \label{fig:dependency_plot1}
\end{figure}

\begin{figure}[htb]
    \centering\includegraphics[width=\linewidth]{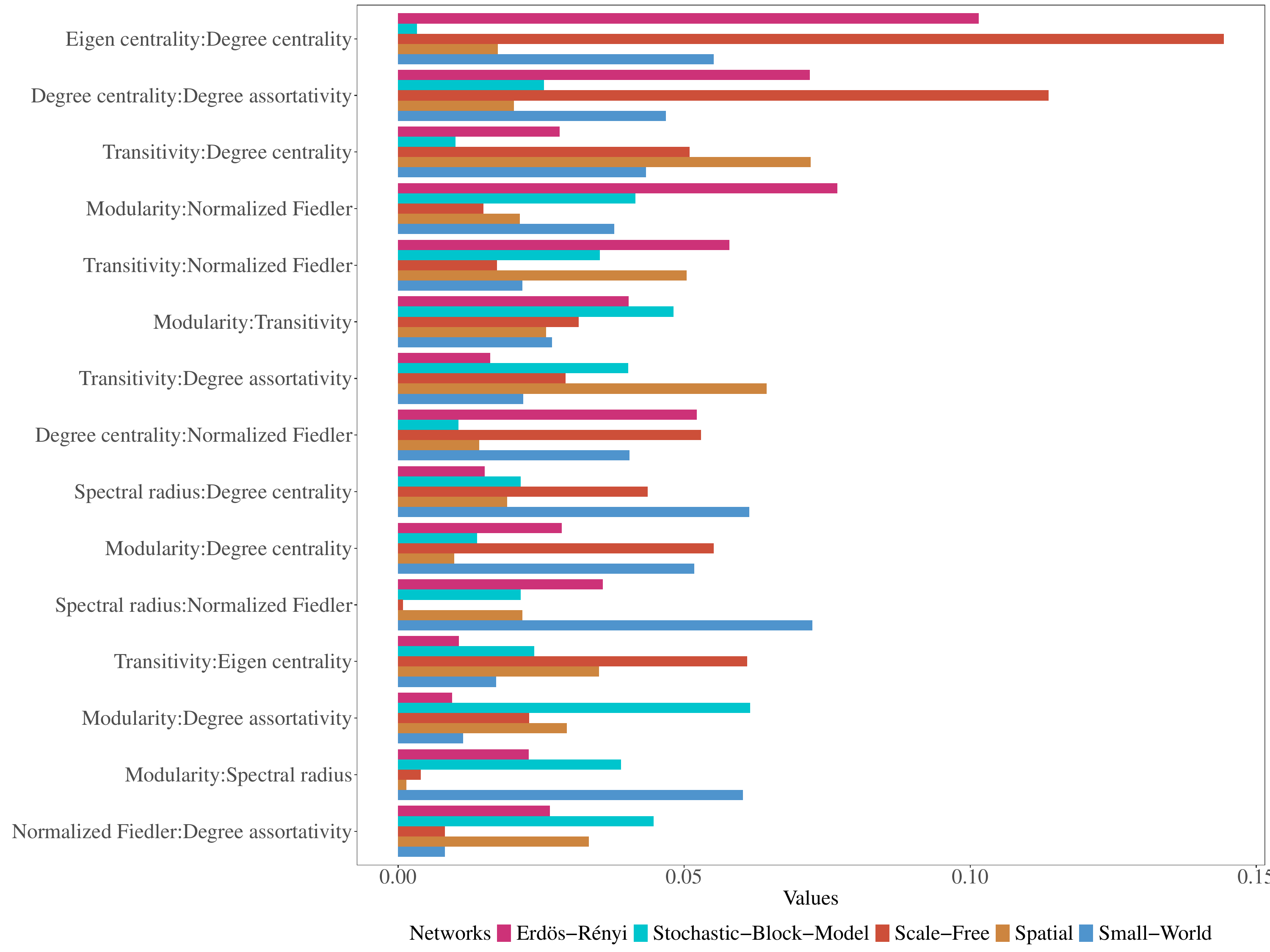}\\
      \caption{Shows the proportion of joint effect variability of two features explained by the  pariwise interactions on Erdös-Rényi, small world, scale free, spatial and stochastic block model predictions.
    The x-axis and y-axis representing the 
 feature values and pairwise feature combinations respectively. 
 The length of the bar associated with the pairwise features typically indicates the strength of the pairwise interaction and how the proportion of the joint effect variability explained by their pairwise interaction influence the models final prediction across the different generative models.}\label{fig:hstats_pairwise} 
\end{figure}

\begin{figure}[htbp]
	\begin{centering}
    \begin{tabular}{l}
        \includegraphics[width=\linewidth]{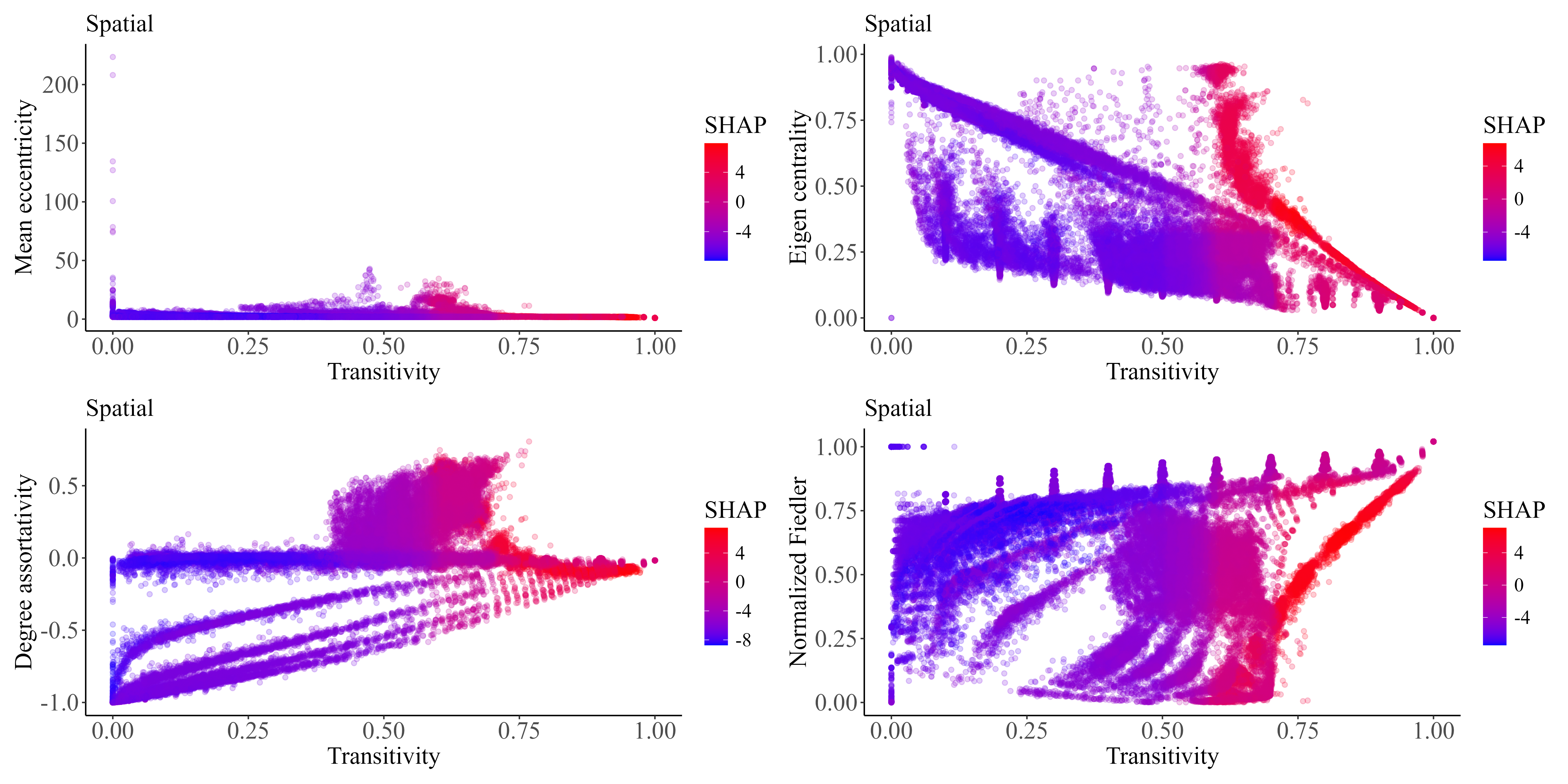} \\[1em]
    \end{tabular}
	\end{centering}
	\caption{SHAP 2D dependency plot showing two-way combined effects of selected variables in predicting the spatial network.
 This plot visualize how the predicted outcome of the spatial network depends on the values of two input features simultaneously.
 The plot represents the dependency between transitivity (on the x-axis) pairing with either mean eccentricity, or eigen centrality, or degree centrality, or degree assortativity, or modularity, or spectral radius in predicting the spatial network with the other features on the y-axis.
 The color or shading of the plot (manually set) indicates the predicted outcome or class probability for the spatial network.
 Overall, this plot shows at which features values the combined effect of the two features existing simultaneously positively, or adversely affect this generative models class prediction.
 It is worth noting that the gaps in this plot signify the absence of simulated networks for specific combinations of feature values, owing to the stochastic nature of the network synthesis and the use of an unknown parameter space beyond our control.}
 \label{fig:2Dplot_sp}
\end{figure}

\subsection{Main and interaction effects} 
\label{sub:feat_interaction}
For all generative models, a substantial proportion of prediction variability remains unexplained by the sum of main effects $H^{2}$ alone (Table.~\ref{fig:classfic}).
We proceed to describe the main network features influencing the prediction of the generative models, and the strength, degree, and form of their interactions with other features in predicting the generative models.
  
\subsubsection{Spatial}
Our investigation unveils transitivity as the strongest positive predictor for spatial networks on both global and local scales (Figs. ~\ref{fig:varimp}(b), ~\ref{fig:dependency_plot1},~S3, ~S4 \& ~S5).
Higher eigen-centrality values, a positive degree assortativity coefficient, and a smaller mean eccentricity and mean path length are additional features that play key roles in predicting this class (Fig. ~\ref{fig:varimp}(b), ~\ref{fig:dependency_plot1} \& Figs.~S3, ~S4 \& ~S5).

Spatial networks in terms of overall feature interactions demonstrate robust interactions associated with transitivity,  normalized Fiedler, degree assortativity coefficient, eigen centrality, and degree centrality (Fig.~\ref{fig:hstats_overall_threeway}(a)).
These features account for over 30\% proportion of prediction variability explained by interactions on them (Fig.~\ref{fig:hstats_overall_threeway}(a)).
Additionally, spatial networks exhibit over 6\% prediction variability from pairwise interactions between transitivity and degree centrality, transitivity and degree assortativity coefficient, and transitivity and eigen centrality (Fig.~\ref{fig:hstats_pairwise}).

Regarding the form of the pairwise interaction, though the pairwise interaction between mean eccentricity and other features predicts this class, the most significant pairwise interaction effects revolves around transitivity values of approximately 0.7 (Fig.~\ref{fig:2Dplot_sp}).
{Regarding the form of the pairwise interaction, the most significant pairwise interaction effects revolves around transitivity values of approximately 0.7 (Fig.~\ref{fig:2Dplot_sp}).}
At this threshold, transitivity shows a strong positive pairwise interaction with other network features (Fig.~\ref{fig:2Dplot_sp}).
Notably, transitivity values at this threshold exhibit positive pairwise interactions with both eigen-centrality values and degree centrality values below 0.5 (Fig.~\ref{fig:2Dplot_sp}).
A positive pairwise interaction is also observed at this 0.7 transitivity threshold alongside mean eccentricity values close to zero (Fig.~\ref{fig:2Dplot_sp}). 
Furthermore, there is a positive pairwise interaction between either modularity or spectral radius and the 0.7 transitivity threshold value (Fig.~\ref{fig:2Dplot_sp}). 
 Moreover, the degree assortativity coefficient values have a strong pairwise interaction with either 
  transitivity or mean eccentricity in predicting spatial networks (Fig.~\ref{fig:hstats_pairwise}).

\subsubsection{Small-World}
The spectral radius stands out as the strongest predictor of this class globally and locally (Fig. ~\ref{fig:varimp}(b), Fig. ~\ref{fig:dependency_plot1}, Figs.~S3, ~S4 \& ~S5).
Generally, small values of the spectral radius predict this class positively, while high spectral radius does not (Fig. ~\ref{fig:varimp}(b), Fig. ~\ref{fig:dependency_plot1}, Figs.~S3, ~S4 \& ~S5).

On the other hand, a high degree centrality, mean degree, and negative degree assortativity coefficient values predict this network negatively.
However, relatively higher transitivity values and smaller values of mean path length positively predict this network (Fig.~\ref{fig:varimp}(b), Fig. ~\ref{fig:dependency_plot1}, Figs.~S3, ~S4 \& ~S5).

In terms of overall interactions in the small-world network, key interactions involve the spectral radius, mean path length, minimum cut,  normalized Fiedler, modularity, transitivity, degree centrality, eigen centrality, and degree assortativity coefficient, explaining approximately 45\% of the proportion of prediction variability  (Fig.~\ref{fig:hstats_overall_threeway}(a)).
Small-world networks display over 6\% prediction variability from pairwise interactions between spectral radius and degree centrality, spectral radius and normalized Fiedler, and spectral radius and modularity  (Fig.~\ref{fig:hstats_pairwise}).

 The form of pairwise interaction shows that when the spectral radius reaches the threshold of approximately 50 or above, it interacts with degree centrality, degree assortativity, transitivity, and modularity revealing a negative pairwise relationship in predicting this class (Fig.~\ref{fig:sf_er_sw_2d_plot}(c)). 
Furthermore, a spectral radius of approximately 50 or higher, has a positive pairwise interaction with small values of mean path length in predicting the small-world network (Fig.~\ref{fig:sf_er_sw_2d_plot}(c)). 
However, we observe a different facet of the spectral radius, wherein its association with higher values of the mean degree negatively influences the small-world prediction (Fig.~\ref{fig:sf_er_sw_2d_plot}(c)).
These pairwise interactions associated with spectral radius and other network features, collectively form a complex relationship that decreases the likelihood of networks being classified as small-world (Fig.~\ref{fig:sf_er_sw_2d_plot}(c)).
We note here that, a high spectral radius interacts with the other features (i.e.,degree centrality,
degree assortativity, transitivity, and modularity) to negatively predict this class, while a small spectral radius interacts with the features to positively predict this class (Fig.~\ref{fig:sf_er_sw_2d_plot}(c)). 
There are other pairwise interactions between either degree centrality and other features like modularity, eigen centrality, normalized Fiedler, transitivity, degree assortativity in predicting the small-world network apart from pairwise interactions between spectral radius and other features (Fig.~\ref{fig:hstats_pairwise}).

While three-way interactions are weaker than pairwise interactions, they remain significant for the generative models (Fig.~\ref{fig:hstats_overall_threeway}(b)). 
Overall, investigating the main feature effects and interactions is important for predicting the generative models as it reveals various feature dependencies and interplay.
These insights aid in identifying influential feature combinations, improving classification accuracy, and the ability to model real-world networks more effectively.

\bigskip
For results on \emph{Erdös-Rényi}, \emph{stochastic-block-model}, and \emph{scale-free} networks, refer to "Additional Results" subsections of the  Supplementary Material.

\subsection{Model application} 
\label{sub: Model application}
To test the utility of our pipeline, we applied our classification model to network data of the western United States power grid \cite{watts1998a}.
Watts and Strogatz \cite{watts1998a} determined the western United States power grid network to be a small-world network, which
our classification model effectively predicted.
Furthermore, our classification model effectively captured the
power-law (scale-free) degree distribution associated with macromolecular networks such as protein and metabolic networks \cite{posfai2016network,schellenberger2010bigg}, biological gene networks and the distribution of the sub-webs within ecological food webs \cite{melian2004food,barab1999a,almaas2007scale}.
In addition, our model predicted that the dynamics associated with badger social organization are spatially determined \cite{weber2013badger}.
The above predictions and the data upon which they are based are available at 
\href{https://github.com/araimacarol/Empirical-Network-Classifier}
{https://github.com/araimacarol/Empirical-Network-Classifier}.

Therefore, our classification pipeline demonstrates application and utility in interpreting and understanding complex structures from network data across diverse domains.
We have attempted to highlight the potential of this approach for future research, for example in exploring the classification of empirical animal social networks.
Moreover, we have created an interactive R-Shiny app accessible online via \href{https://araima.shinyapps.io/Empirical-Network-Classifier}
{https://araima.shinyapps.io/Empirical-Network-Classifier}.
This app allows users to upload new network data and predict their generative model class using graph features employed during model training. 
Additionally, the app presents summaries of estimated graph features, and incorporates SHAP model interpretation \cite{lundberg2017unified,marcilio2020explanations}, and Friedman and Popescu’s H-statistics \cite{friedman2008predictive} for visualizing feature importance, interactions, and quantifying the degree and strength of the interactions.
This can aid users and in understanding key features influencing the prediction of the generative models class by the model and the relevance of their interactions on the model's prediction.

 \section{Discussion} 
 \label{sec:Discussion}
Our machine learning approach provides a powerful method to accurately predict generative models for empirical networks and interpret model outputs.
Additionally, we leveraged advances in machine learning interpretation tools like Shapley Additive Explanations (SHAP) \cite{marcilio2020explanations,lundberg2017unified} and Friedman and Popescu’s H-statistics \cite{friedman2008predictive}  to explore  feature effects and their interactions in distinguishing between generative models.
This reveals varying feature sets associated with each model and enhances the robustness of predictions for generative models.

Our findings highlight the absence of consistent patterns associated with features across models, suggesting diverse feature requirements for accurate prediction for different models. 
Likewise, the interaction among distinctive features, including spectral measures such as the (Normalized) Fiedler value, spectral radius, and eigen centrality, alongside other network properties like transitivity, degree assortativity coefficient, modularity, and degree centrality (see Subsection \ref{sub:notation_and_description_of_network_features}), differs across various generative models. 
Ikehara \etal\ \cite{ikehara2017characterizing}, and  Canning \etal\ \cite{canning2017network},
have classified networks using network features like clustering coefficient (transitivity), mean degree, and degree assortativity coefficient, which are subsets of our feature set. 
Our work advances these studies by incorporating a diverse array of theoretical network models, integrating spectral measures into the feature set, and exploring a broader parameter range and combinations.
Moreover, our study highlights the features' impact and most importantly their interactions, which are less understood in the context of predictive modeling and classification tasks.

We have demonstrated the effectiveness of our pipeline, achieving nearly 100\% prediction and classification accuracy alongside an impressive AUC (Area Under the ROC Curve) score of nearly 1.0 across $10$-fold cross-validation samples. 
Our classification accuracy outperforms the 94.2\% of Canning \etal\ \cite{canning2017network}, 95\% of Barnett \etal\ \cite{barnett2019a}, Ikehara \etal\ \cite{ikehara2017characterizing},
and between 55.7\%-91.4\% that Nagy \etal\ had on their binary and multi-class classification model \cite{nagy2022a}. 
Our approach identifies distinct features distinguishing the prediction and/or classification of diverse networks, aligning closely with previous research
\cite{canning2017network,ikehara2017characterizing,nagy2022a}. 
However, the diversity of our approach makes it suitable for classification tasks since it enables a wide spectrum of structural properties observed in real-world networks such as sparsity, small-world characteristics, clustering (transitivity), path length, degree distribution, and sub-graphs to be captured and estimated.

In addition, incorporating spatial networks enhances our approach \cite{janssen2012model}. 
The inclusion of spatial networks not only adds robustness to simulating large networks but also reproduces observed network properties such as sparsity, small-world characteristics, and clustering.
Moreover, we provide compelling evidence that a sufficiently rich collection of structural features, particularly spectral measures like the (Normalized) Fiedler value, eigenvector centrality, and spectral radius, capture important structural heterogeneity in networks \cite{wills2020metrics}. 
This allows our model to predict the generative model for network organization across different domains, sharing similar structural properties.
The exceptional performance of our pipeline negates the need for complex features like network motifs or graphlets employed in previous studies, which can incur considerable computational complexity \cite{janssen2012model,ikehara2017characterizing,nagy2022a,canning2017network}.
  
Our findings across various generative models are consistent with well-established characteristics of theoretical networks \cite{shirley2005a,keeling2011networks,wills2020metrics,erdos1960a,keeling2005networks,newman2003,ames2011a}. 
For instance, Erdős-Rényi networks, known for their randomness and low clustering \cite{shirley2005a,keeling2011networks,wills2020metrics,erdos1960a,keeling2005networks,newman2003}, can also be accurately identified by emphasizing the often-overlooked spectral property, the normalized Fiedler.

Similarly, predictors such as degree centrality, eigen centrality, and negative degree assortativity coefficient emerge as key indicators for scale-free networks, consistent with prior research \cite{newman2003,barab1999a,shirley2005a,keeling2005networks,keeling2011networks}. 
In spatial networks, primary predictors include transitivity, mean eccentricity, and diameter aligning well with prior studies \cite{yiu2004clustering,newman2003,janssen2012model}. 
Moreover, we have identified that spatial networks are associated with high eigenvector centrality, assortativity mixing, and small normalized Fiedler due to its sparsity. 
Similarly, modularity, mean path length, and positive degree assortativity coefficient  are relevant in predicting the stochastic block models, aligning with findings from previous research \cite{newman2003,wills2020metrics}. 
Our study also identified spectral radius as a predictor of the small-world network, an insight not initially known. Moreover, we found that the small-world network, and scale free networks are disassortative, whiles spatial and stochastic block models are mostly assortative.

Furthermore, we identified threshold values where interactions among network features influence predictions. 
Notably, key predictors like transitivity, spectral radius, modularity, and normalized Fiedler remained important across a large range of training input over all simulations, consistently appearing among the top predictors for spatial, small world, stochastic block models, and Erdős-Rényi networks respectively.

Our study highlights the importance of considering network features including spectral properties like the spectral radius, normalized Fiedler, as well as node relation network properties like the degree assortativity coefficient, alongside tailored predictors for each generative model.
This approach facilitates effective network classification and enhances our understanding of the complex structure and dynamics of networks, indicating the robustness, community structure, and potential pathways for processes like  disease spread \cite{fountain2023spectral}.
Overall, our analysis identified that features such as normalized Fiedler value, modularity, transitivity, degree centrality, spectral radius, degree assortativity coefficient, and eigen centrality are the most relevant features for predicting the generative models.

While our study provides valuable insights, it is important to acknowledge certain limitations. Notably, we did not consider generative models such as 2K, Forest Fire, Kronecker graphs, and Exponential Random Graph Models (ERGMs) \cite{krivitsky2014separable,leskovec2010kronecker,leskovec2006sampling,hunter2008ergm}. 
Although these models may offer advantages in specific contexts like dynamic or evolving structures and
nested hierarchical organizations found in certain networks \cite{nagy2022a,wills2020metrics,leskovec2010kronecker,hunter2008ergm,janssen2012model}, they may not be suitable for estimating empirical networks. 
For instance, Kronecker graphs suffer from scalability, model specification challenges, and lack of realism and interpretability \cite{leskovec2006sampling,leskovec2010kronecker}. Similarly, ERGMs face scalability issues, model degeneracy, and computational complexity when applied to large networks \cite{hunter2008ergm}. Moreover, ERGMs, 2K models, and Forest Fire models all require significant computational time, and they produce complex parameter estimates that are difficult to interpret \cite{hunter2008ergm}. 
Conversely, the generative models we have chosen like Erdös-Rényi, stochastic block model, scale-free, small world, and spatial, although some may be seem simplistic like the Erdös-Rényi networks, offer undeniable relevance in simulating real world networks. While these models may not also capture certain properties as comprehensively as the excluded ones, they excel in reproducing fundamental network characteristics, and offer better interpretability, computational efficiency, and applicability to various network types, enabling meaningful insights into network organization and dynamics \cite{shirley2005a,watts1998a,barab1999a,newman20102010networks,keeling2005networks,keeling2011networks}.

Ongoing future research aims to develop a hybrid generative model, such as a spatial scale-free weighted model, capable of capturing a diverse range of properties from various networks. Additionally, potential future endeavors could delve into integrating the 2K, Forest Fire, Kronecker graphs, and Exponential Random Graph Models (ERGMs) \cite{krivitsky2014separable,leskovec2010kronecker,leskovec2006sampling,hunter2008ergm}, to deepen our comprehension of network formation and dynamics across diverse domains.

\maketitle

\section{Conclusion}
In conclusion, our method effectively identifies distinct features and their interactions in distinguishing the generative models. 
We achieved this by using novel interpretable machine learning tools like SHAP and Friedman and Popescu’s H-statistics \cite{friedman2008predictive,marcilio2020explanations,lundberg2017unified}.
Additionally, we utilized generative models as labels, enhancing our feature sets with spectral measures, simulating a substantial set of networks, and significantly expanding the parameter space over previous studies.
Our results achieved an outstanding classification accuracy, nearly 100\%, surpassing the performance of previous studies.
 Our pipeline, together with interpretable machine learning approaches, forms a potent tool for the identification of significant feature dependence and interactions, and enhances our understanding of the underlying generative models. 
This in-depth understanding contributes to advancements in network classification and analysis, ultimately facilitating more accurate and effective modeling of complex systems.
We hope that this work will provide both a useful tool and important insights into the analysis of biological and ecological networks. 
.

\section{Acknowledgements}
This project was supported by an Australian Research Council Discovery Project Grant (DP190102020).

\section{Data statement} 
All the data and code to perform the analysis can be found at \href{https://github.com/araimacarol/Empirical-Network-Classifier}{https://github.com/araimacarol/Empirical-Network-Classifier}

 \newpage
\printbibliography

\clearpage 

\appendix
\section*{Supplementary material from "Leveraging advances in machine learning for the robust classification and interpretation of networks.”}
\renewcommand{\thefigure}{S\arabic{figure}} 

\setcounter{section}{1} 
\renewcommand{\thesection}{\arabic{section}} 

\subsection{Definition of Graph Features}

First, we differentiate the local and global metrics used:

\begin{itemize}
    \item  \emph{Local metrics} describe individual nodes in a network \cite{ames2011a,newman2003,newman20102010networks,newman2013spectral,freeman2002centrality,wasserman1994social}. 
    Local features include centrality measures, which are used to assess the importance, influence, or dominance of nodes within a network \cite{wasserman1994social,newman2003,newman20102010networks,barabasi2016network,ikehara2017characterizing,janssen2012model}. 
    Additionally, these measures help identify key nodes influencing the overall network's structure, detect communities or clusters, and assess the interaction of specific nodes on their neighbor's connectivity, and dynamics \cite{wasserman1994social,newman2003,newman20102010networks,barabasi2016network,ikehara2017characterizing,janssen2012model}. 
    Examples include degree centrality, local clustering coefficient, node degree centrality, node betweenness centrality, and node closeness centrality \cite{ames2011a,keeling2011networks,wasserman1994social,newman2003,newman20102010networks,barabasi2016network,janssen2012model}.
    \item 
    \emph{Global metrics} describe features of the network as a whole \cite{ames2011a,shirley2005a,wasserman1994social,newman2003,newman20102010networks,barabasi2016network,freeman2002centrality}. 
    They are useful for understanding the overall connectivity, resilience, efficiency, and organization of the network, identifying patterns or anomalies at the network level, and comparing networks with different structures or characteristics. \cite{ames2011a,wasserman1994social,newman2003,newman20102010networks,barabasi2016network,ikehara2017characterizing,janssen2012model,wills2020metrics}. 
    Examples include network density, mean degree, minimum cut size, mean path length, mean degree, diameter, modularity, degree assortativity coefficient, spectral radius, and (Normalized) Fiedler \cite{freeman2002centrality,shirley2005a,wills2020metrics,newman2003,newman2013spectral,ames2011a,newman20102010networks,keeling2011networks}. 
    In this study, we incorporate global metrics, alongside the global index or formulation of local metrics, as features for graph classification
\end{itemize}

\paragraph{Degree centrality:} 
\label{ssub:degree_centrality}

 Degree centrality measures how the degrees of nodes in a graph differ from the degree of the most central node \cite{freeman2002centrality,wasserman1994social}. 
 It is commonly employed to evaluate the concentration of the network structure around a few highly connected nodes \cite{freeman2002centrality,wasserman1994social}.
 Nodes with high degrees assume crucial central roles in graphs and are important for the overall network functionality \cite{kajdanowicz2016using}. 
 We define the degree centrality of graph $G$ below using Freeman's general formula \cite{metcalf2016cybersecurity}.

 \[
	d_{c}(G) =
 \frac{1}{V_{norm}}\sum_{v \in G}\left[deg(v')-deg(v) \right],
\] 
where $V_{norm}=(|v|-1)(|v|-2)$, and $v'$ is the most central node (node with the highest degree) \cite{metcalf2016cybersecurity}.
We take the graph-level degree centrality index by summing the degree centrality scores of all nodes and then normalizing them by dividing by the theoretical maximum (the most centralized graph with the same
number of nodes as the underlying graph).

\paragraph{Closeness Centrality:} 
\label{ssub:closeness_centrality}
Closeness centrality is defined as the average shortest path length from a node to every other node in the network (a measure of the number of steps to access every other vertex from a given vertex) \cite{kajdanowicz2016using,freeman2002centrality,wasserman1994social}.
We define the closeness centrality of vertex $v_i$ based on Freeman's approach as

 \[
	c_{c}(v_i) =
 \frac{1}{\sum_{v_{i}\neq v_{j}} d({{v_i}{v_j}})},
\] 
where $d(v_{i},v_{j})$ correspond to the Euclidean distance between node $v_{i}$ and $v_{j}$ \cite{freeman2002centrality}. 
A global summary is obtained by averaging the closeness centrality of all nodes in a graph.
The global gives the average distance or closeness of the entire graph. 
It represents the average ``closeness" of all nodes to each other in the graph, which provides a global measure of how well-connected the graph is overall.

\paragraph{Betweenness centrality:} 
\label{ssub:betweenness_centrality}

 Betweenness centrality (here we refer to node betweenness) is defined as the number of geodesic distances (shortest path) traversing a node.
\cite{kajdanowicz2016using,freeman2002centrality,wasserman1994social}.
The betweenness centrality of a node $v_i$ is defined as

\[
	b_{c}(v_i) =
{\sum_{v_{i}\neq v_{j}\neq v_{k}}} 
\frac{g({v_j}{v_i}{v_k})}{g({v_j}{v_k})},
\] 
where $g({v_j}{v_k})$ is the total number of shortest path between node $v_j$ and $v_k$ respectively, and $g({v_j}{v_i}{v_k})$ is the number of those shortest paths traversing node $v_i$ \cite{freeman2002centrality}.
Like the degree centrality, we take the graph-level betweenness centrality index by summing the betweenness centrality scores of all nodes and then normalizing them by the theoretical maximum.

\paragraph{Eigenvector centrality:} 
\label{ssub:eigen_centrality}

This measure centralizes a graph according to the eigenvector centrality of nodes \cite{freeman2002centrality}. 
Eigen centrality considers not only the number of direct connections a node has (like degree centrality) but also the ``influence'' of the neighbouring nodes in the graph \cite{bonner2016efficient}.
It assumes that a node has high ``influence'' (high eigenvector score) if it is connected to other important nodes with high eigenvector score \cite{bonner2016efficient}.
The eigenvector centrality $e_{c}$ of node $(v_i)$ is defined as

\[
	e_{c}(v_i) =
 \frac{1}{\alpha}\sum_{v_j}A_{{v_i}v{_j}}e_{c}(v_j),
\] 
where $A_{{v_i}v{_j}}$ is the adjacency matrix of the graph, and $\alpha$ is a proportionality constant \cite{das2023sentiment}. 
We take the graph-level eigenvector centrality index as a global metric by summing the eigenvector centrality scores of all nodes and then normalizing them by the theoretical maximum.

\paragraph{Mean eccentricity:} 
\label{ssub:mean_eccentricity}
The \emph{eccentricity} of a node $v_i$ is the maximum distance from that node to any other node in the graph \cite{li2012a,harary2018graph,bondy1982graph}. This is defined as

\[
ecc (v_i) = \max \{d(v_i,v_j): v_i,v_j \in V\} , 
\]
where the distance $d(v_i,v_j)$ is the length of the shortest path traversed from node $v_i$ to $v_j$ \cite{li2012a,harary2018graph}.
The eccentricity consists of the score of each given node in a graph.
We consider the mean of the eccentricity as a global metric of a graph.
Therefore, the \emph{mean eccentricity} of a graph is the average eccentricity of all nodes in a connected graph \cite{li2012a,harary2018graph,bondy1982graph}, and this is given as

\[
\hat{E} = \frac{1}{n}\sum^{n}_{i=1}ecc(v_i).
\]
This measure indicates how far apart the nodes are from each other on average in a graph.
A smaller mean eccentricity indicates that the graph is more tightly connected (clustered), with nodes being closer to each other on average \cite{li2012a,harary2018graph,bondy1982graph}.

\paragraph{Diameter:} 
\label{ssub:Diameter}
The \emph{diameter} of a graph is defined as the maximum eccentricity. 
The diameter represents the longest shortest path between any pair of nodes in the graph.
In other words, it is the maximum distance between any two nodes in the graph. 
This is defined as

\[
D = \max_{1\leq i\leq n} \{ecc(v_i)\}.
\]
The diameter, which is a global graph metric provides a measure of the overall size of the graph and how spread out its nodes are.

\paragraph{Radius:} 
\label{ssub:Radius}
The \emph{radius} of a graph is also defined as the minimum eccentricity in a graph 
This metric is represented as
\[
R = \min_{1\leq i\leq n} \{ecc(v_i)\}.
\]
This global metric represents the shortest maximum distance from any node to all other nodes, indicating the ``center" of the graph.

\paragraph{Mean path length :} 
\label{ssub:mean path length}
The \emph{mean path length} (or mean geodesic distance) of a graph measures the average number of steps traverse from a node $v_i$ to each other node in a graph \cite{li2012a,harary2018graph,keeling2011networks,newman20102010networks}. 
This global measure is defined as

\[
\hat{L} = \frac{1}{n(n-1)}\sum^{n}_{i=1}\sum^{n}_{j \neq i}d(v_i,v_j).
\]
Unlike mean eccentricity, diameter, and radius, mean path length considers every possible pair of nodes in the graph and averages their shortest path lengths, giving insight into the graph's connectivity and how efficiently information can occur between vertices \cite{harary2018graph}.

\paragraph{Clustering coefficient (Transitivity):} 
\label{ssub:clustering_coefficient}
The \emph{local clustering coefficient} is defined as the proportion of a node's neighbors that are also neighbors of each other  \cite{wills2020metrics}. 
The clustering coefficient of a node $v_{i}$ can be written
\[
c(v_{i}) = \frac{\sigma(v_{i})}  {\varphi(v_{i})},
\]
where $\sigma(v_{i})$ is the number of fully connected triples or triangles (complete graph with three nodes) of node $v_{i}$ and $\varphi(v_{i})$ denotes the number of pairs of neighbours that are adjacent to node $v_{i}$ \cite{li2012a}.
The \emph{global clustering coefficient} (transitivity) of a graph is the average measure over all nodes in the graph \cite{li2012a},
\[
c(G) = \frac{1}  {n} \sum_{i=1}^{n} c(v_{i}).
\]
The clustering coefficient $c(G)$ can take on values close to zero for networks such as chemical compounds with very few triangles \cite{li2012a}.
The global clustering coefficient or transitivity quantifies the extent to which nodes in a network tend to cluster together \cite{li2012a}. 

\paragraph{Degree assortativity coefficient:} 
\label{ssub:degree_assortativity_coefficient}

This is a measure of the level of homophily (tendency of nodes to connect with other nodes with similar attributes or characteristics) with respect to some node labeling or values \cite{newman2003,carnegie2018a}. 
Although there are important node attributes such as sex, race, and socio-economic status, social network scientists preferably quantify this measure in terms of node degree \cite{newman2003,carnegie2018a}.
Newman \etal\ described this global metric as
\[
	r =
 \frac{1}{\sigma^{2}_{q}}\sum_{v_jv_k}v_jv_k\left[e_{v_jv_k}-q_{v_j}q_{v_k} \right],
\] 
where
$q_{i}$ = $\sum_{j}$ $e_{v_iv_j}$ for undirected graphs and $e_{ij}$ is the fraction of edges connecting nodes $v_i$ and $v_j$ \cite{newman2003}. This means that
\[r =
 \frac{1}{\sigma_{o}\sigma_{i}}\sum_{v_jv_k}v_jv_k\left[e_{v_jv_k}-q^{o}_{v_j}q^{i}_{v_k} \right],
\] 
where $\sigma_{o}$, $\sigma_{i}$, $\sigma_{q}$ are the standard deviations of $q$, $q^{o}$, and $q^{i}$ \cite{newman2003}.
 A Positive degree assortativity coefficient is also known as \emph{assortativity} mixing, and negative degree assortativity coefficient is also known as \emph{dissassortative} mixing \cite{newman2003}.

\paragraph{Minimum, Maximum and Mean degree:}
\label{ssub:min_degree}
Using the standard definitions \cite{bondy1982graph}, the minimum degree of a graph $G$, denoted as $\delta(G)$, is defined as the minimum number of all the node degrees of graph $G$:
\[
\delta(G) = \min\{\deg(v) \mid v \in V(G).\}
\]
Similarly, the maximum degree of a graph $G$ is the maximum number of all the node degrees of graph $G$:
\[
\Delta(G) = \max\{\deg(v) \mid v \in V(G)\}
\]
The mean degree of a graph $G$ denoted as $\hat{d} (G)$ is defined as the average number of all the node degrees of graph $G$:
\[
\hat{d} (G) = \frac{1}{|V|}\sum_{v\in V(G)}\deg(v)
\]

\paragraph{Modularity:} 
\label{ssub:modularity}

This is a global measure of the strength of the division of a graph into modules or sub-structures \cite{newman2013spectral,sah2017a,newman2003}. 
Networks with higher modularity scores will have more connections within the same community and less between communities.
Modularity can be defined as
\[
	Q = \sum_{u=1}^{U}\left[\frac{q_{u}^{w}}{q}- \left(\frac{q_{u}}{q}\right)^{2}\right],
\]
 where $q_{u}$ is the total number of edges in sub-structure $u$ with $q_{u}^{w}$ denoting the specific edges in the sub-structure and $q$ represent the total number of edges in the network \cite{sah2017a}.

\paragraph{Minimum Cut size:} 
\label{ssub:minimum_cut}
A \emph{minimum cut size} of a graph is the minimum total number of edges required to separate the network into more connected components (at least two components) and is a coarse measure of the connectedness of the network \cite{stoer1997simple,seary2003spectral}.
The minimum cut problem aims to find a partition of the vertex set $V$ into two disjoint sets 
$\textsc{A}$ and $\textsc{B}$ such that the number of edges between the two sets, known as the cut size, is minimized.
Let $\eta (S)$ denote the cut size of the partition $S$, where 
$S$ $\subseteq V$ \cite{stoer1997simple,seary2003spectral}.

\paragraph{Graph energy:} 
\label{ssub:Graph energy}
The \emph{graph energy} is defined as the squared sum of the absolute values of all eigenvalues of the adjacency matrix $A$ \cite{li2012a,harary2018graph}. 
This is given as 

\[
 E_G = \sum^{n}_{j=1}(\lambda^{A}_{j})^2
\]

\subsection{Additional Results}

These are the additional results on the generative models from Section 3.1 of the main document. 
Results on \emph{spatial} and \emph{small world} networks can be found in the Subsections (3.1.1-3.1.2) of the main document. 

\subsubsection{Erdös-Rényi}
We discovered that the normalized Fiedler value displays the strongest positive correlation with Erdös-Rényi network prediction, both globally and locally (Figs.~3(a), 4, ~S4 \& ~S5).
However,higher values of transitivity, modularity, and degree centrality collectively have an adverse impact on Erdös-Rényi prediction, while lower lower values of these features is indicative of the Erdös-Rényi network  (Figs.~3(a), 4, ~S4 \& ~S5).
These findings on a global scale can be visualized by the SHAP dependency and SHAP feature effect or importance plots (Figs.~3(a), 4, ~S4 \& ~S5). 
On a local scale (an instance of prediction), examining individual data instances reveals that all these features exhibit a negative correlation in predicting Erdös-Rényi networks (Fig.~\ref{fig:waterfall}). 
Nevertheless, these SHAP local interpretation unlike the global ones pertains to individual data instances, and their generalizability to universal trends may be limited.

Regarding feature interactions in this class, the majority of the overall interaction effects are linked to normalized Fiedler, modularity, transitivity, and degree centrality (Fig.~\ref{fig:hstats_overall_threeway}(a)).
These features collectively contribute to over 50\%  of the proportion of prediction variability explained by interactions on these features (Fig.~\ref{fig:hstats_overall_threeway}(a)).
Additionally, over  25\% of the proportion of the joint effect variability comes from pairwise interactions between normalized Fiedler and other features like degree centrality, transitivity, spectral radius, degree assortativity coefficient, and modularity (Fig.5). 
 However, the joint effect variability of over 20\% is observed from the pairwise interaction effect between degree centrality and eigen centrality, degree centrality and degree assortativity coefficient, and between transitivity and modularity in explaining the prediction of Erdös-Rényi network (Fig. 5).

 Employing SHAP 2D (two-dimensional) interaction plots to demonstrate the form of the pairwise interactions for Erdös-Rényi network, shows that
the strongest pairwise interactions for predicting Erdös-Rényi networks are associated with the normalized Fiedler value, roughly 0.75 or higher (Fig.~\ref{fig:sf_er_sw_2d_plot}(b)).
The pairwise interaction of the normalized Fiedler at the 0.75 threshold value and degree centrality at 0.25 or lower positively impacts this class prediction (Fig.~\ref{fig:sf_er_sw_2d_plot}(b)). 
Additionally, the pairwise interaction of normalized Fiedler at this threshold and modularity values below 0.25, contributes positively to Erdös-Rényi predictions (Fig.~\ref{fig:sf_er_sw_2d_plot}(b)). 
A positive pairwise interaction materializes between the normalized Fiedler at the 0.75 threshold value and any level of transitivity in predicting Erdös-Rényi networks (Figs.~S2(b)).
Similarly, a strong positive pairwise interaction is also observed between the normalized Fiedler at the 0.75 threshold value and either the mean degree or spectral radius in predicting this network (Figs.~S2(b)).

\subsubsection{Stochastic Block Model}
 Modularity and positive degree assortativity coefficient emerged as the strongest global and local positive predictor for the stochastic block model (Figs. 3(a), 4, ~\& S5). 
A high graph energy also contribute positively to the prediction of the stochastic block model on both scales, while a higher normalized Fiedler adversely affects this class prediction (Figs. 3(a), 4, ~S3 ~\& S5).
In contrast, smaller mean path length, mean eccentricity, and large transitivity values tend to have adverse effects, both locally and globally on the prediction of this network (Figs. 3(a), 4, ~S3 ~\& S5). 
In terms of overall feature interactions within this class, the stochastic block model demonstrates the significance of modularity, transitivity, normalized Fiedler, mean path length, degree assortativity coefficient, degree centrality, and spectral radius, in explaining about 40\% of the prediction variability (Fig.~\ref{fig:hstats_overall_threeway}(a)).
 
 The pairwise interaction in the stochastic block model
 are mostly associated with modularity and degree assortativity coefficient (Fig. 5).
 About 25\% of joint prediction variability can be attributed to pairwise interactions between modularity and other features like spectral radius, transitivity,normalized Fiedler, and degree assortativity coefficient (Fig. 5).
 Additionally, 15\% joint prediction variability are attributed to the pairwise interactions  between degree assortativity coefficient and other features like degree centrality, normalized Fiedler, and transitivity (Fig. 5).

Exploring the form of pairwise interactions for this class shows that the positive prediction of this class is strongly influenced by modularity, particularly at the threshold values in the range of 0.1 to 0.35 (Figs.~S2(a)). 
Moderate modularity values at this threshold interacts with mean path length and various other features (Figs.~S2(a)).
Notably, positive pairwise interactions occur between these moderate modularity threshold values and mean path length values near zero, while negative interactions happen outside this range (Figs.~S2(a)).
Morever, a positive pairwise interactions are also observed between positive degree assortativity coefficients and the moderate modularity threshold values (Fig.~\ref{fig:sf_er_sw_2d_plot}(a)). 
Similarly, the moderate modularity values exhibit a  positive pairwise interaction with graph energy and transitivity, enhancing this model's class prediction (Fig.~\ref{fig:sf_er_sw_2d_plot}(a)).
Additionally, positive pairwise interactions are noted between the moderate modularity values and spectral radius less than or equal to 750, as well as high modularity values and normalized Fiedler values less than or equal to 0.75 (Fig.~\ref{fig:sf_er_sw_2d_plot}(a)). 

Our results also demonstrate a positive prediction of this class associated with the pairwise interactions between degree assortativity coefficient and other features (Fig.~\ref{fig:sf_er_sw_2d_plot}(a)).
Particularly, the pairwise interaction effect between positive degree assortativity coefficient and high normalized Fiedler, and transitivity, or between  positive degree assortativity coefficient and small closeness centrality and mean path length values predicts this class (Fig.~\ref{fig:sf_er_sw_2d_plot}(a)).  A pairwise interaction between positive degree assortativity and graph energy also predicts this class (Fig.~\ref{fig:sf_er_sw_2d_plot}(a)). 

\subsubsection{Scale-Free}
For the scale-free network, our analyses demonstrate that degree centrality and eigen centrality stand out as the strongest predictors on both global and local scales (Figs. 3(b), 4, ~S3, ~S4 \& ~S5).
 However, the presence of modularity, graph energy, normalized Fiedler value, and positive degree assortativity coefficient values adversely affects the prediction of this class (Figs. 3(b), 4, ~S3, ~S4 \& ~S5).

 In terms of overall feature interactions, scale free networks exhibit strong overall interactions with degree centrality and eigen centrality (Fig.~\ref{fig:hstats_overall_threeway}(a)).
These centrality measures, alongside transitivity, and graph energy, explain approximately 30\% proportion of prediction variability (Fig.~\ref{fig:hstats_overall_threeway}(a)).
 The pairwise interactions involving degree and eigen centrality contribute to about 15\% of the joint prediction variability, while pairwise interactions between degree centrality and degree assortativity coefficient 
  contribute to about 12\% of the joint prediction variability in predicting this class (Fig. 5).
  Also, we note that the pairwise interaction between degree centrality and transitivity or normalized Fiedler, or spectral radius or modularity account for over 5\% of the joint effect prediction variability (Fig. 5). This is also the case for pairwise interaction between eigen centrality ad transitivity (Fig. 5).

Exploring the form of pairwise interaction for this class also shows that the strongest pairwise interaction effects for predicting the scale-free network primarily center around degree centrality and eigen centrality values exceeding approximately 0.2 (Fig.~\ref{fig:sf_er_sw_2d_plot}(d)).
At this threshold value for the centrality measures, a positive pairwise interaction occurs (Fig.~\ref{fig:sf_er_sw_2d_plot}(d)). 
Additionally, the pairwise interaction between negative degree assortativity coefficient and degree centrality values exceeding 0.2 positively predicts this class (Fig.~\ref{fig:sf_er_sw_2d_plot}(d)).
 Degree centrality values at or above 0.2 and graph energy values less than 5000 have a strong positive pairwise interaction in predicting this class (Fig.~\ref{fig:sf_er_sw_2d_plot}(d)).
This positive relationship extends to spectral radius values less than 750, and degree centrality values at or above 0.2 (Fig.~\ref{fig:sf_er_sw_2d_plot}(d)). 
The same dynamics are evident when degree centrality values surpass 0.2 and transitivity values are below 0.75 (Fig.~\ref{fig:sf_er_sw_2d_plot}(d)). 
Furthermore, when modularity values fall below 0.2, and degree centrality values remain at or exceed 0.2, a positive pairwise interaction effect materializes in predicting this network class (Fig.~\ref{fig:sf_er_sw_2d_plot}(d)).

\setcounter{figure}{0} 

\begin{landscape}
\begin{figure*}[b]
    \centering
\begin{tabular}{l}
         (a) \\
\includegraphics[scale=0.3]{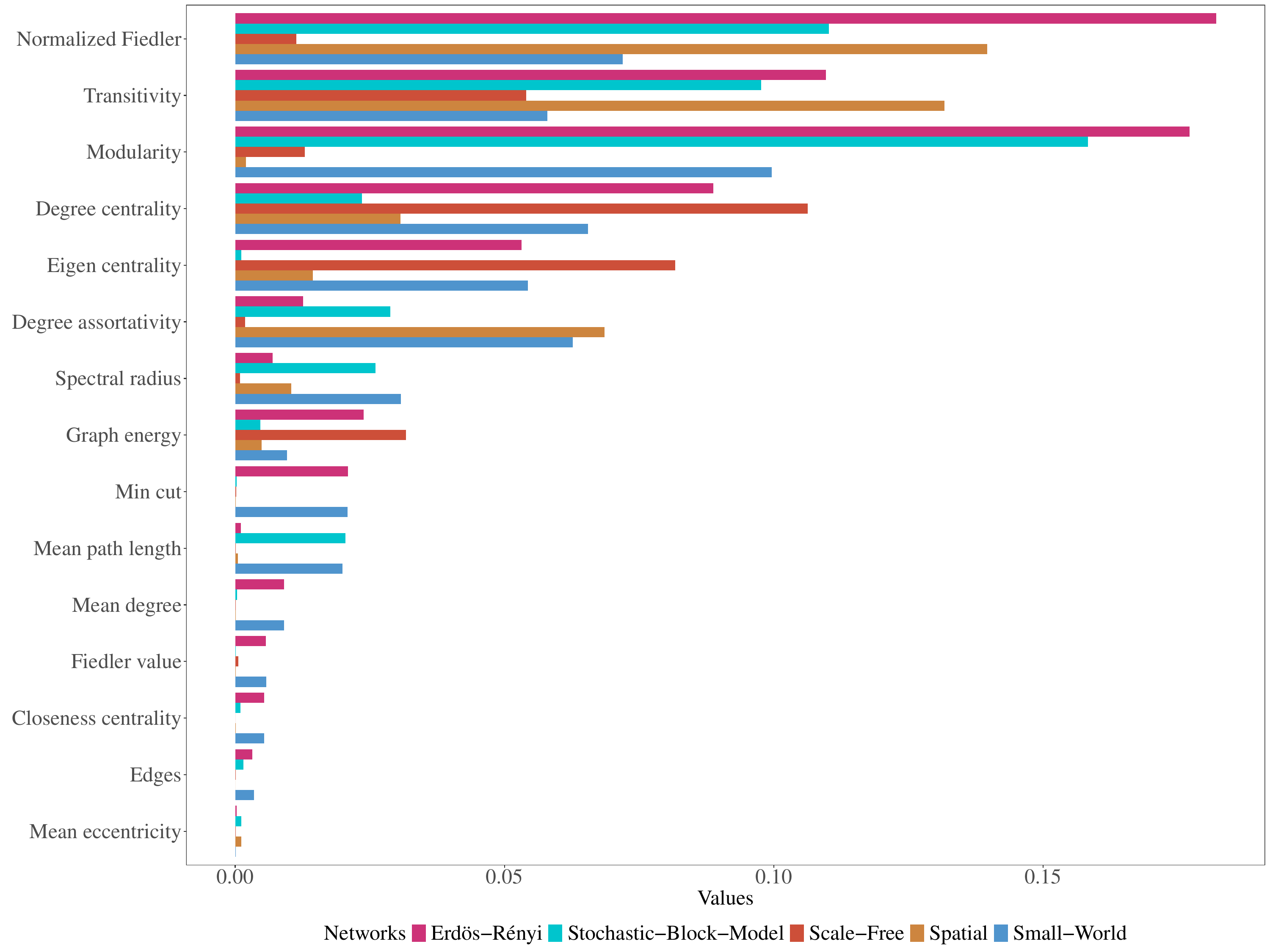}\\
\end{tabular}
\end{figure*}

\end{landscape}

\begin{landscape}
\begin{figure*}[b]
   \centering
\begin{tabular}{l}
         (b) \\
\includegraphics[scale=0.3]{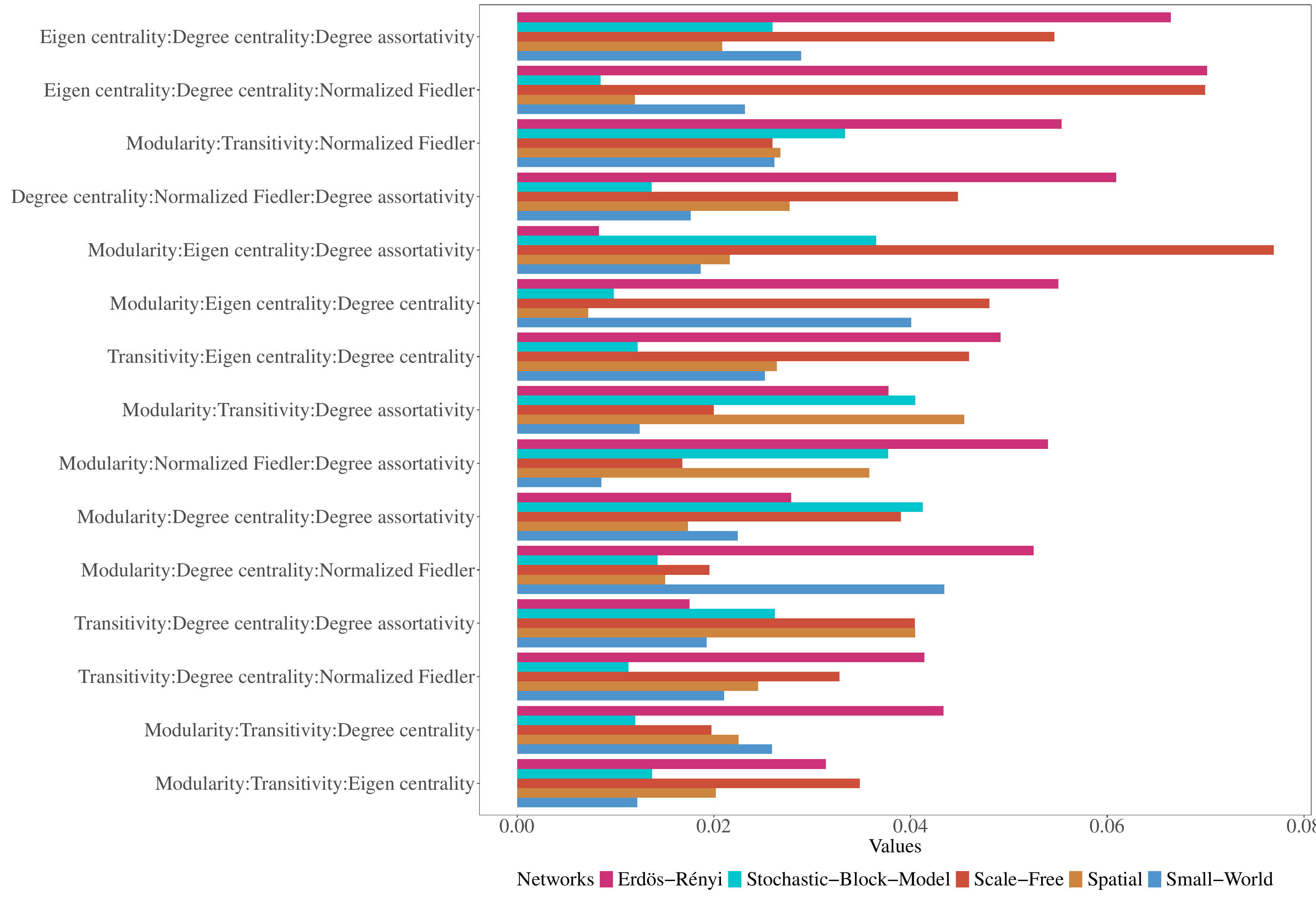}\\
\end{tabular}    
\end{figure*}

\end{landscape}

\begin{figure*}[b]
     \centering
   \caption{This plot shows the proportion of predicted variability for the (a) strongest overall interaction with the main feature effects, and (b) three-way interaction between features on Erdös-Rényi, small world, scale free, spatial, and stochastic block model predictions.
    The x-axis and y-axis represent the 
 feature values and main feature, or three-way feature combinations respectively for the (a) overall interaction and (b) three-way plots. 
 The length of the bar associated with the features on the y-axis typically indicates the strength of the overall or three-way interaction and how these interactions influence the model's prediction across the different generative models.}
\label{fig:hstats_overall_threeway} 
\end{figure*}

 \begin{landscape}
\begin{figure*}[b]
\centering
\begin{tabular}{l}
         (a) \\
\includegraphics[scale=0.4]{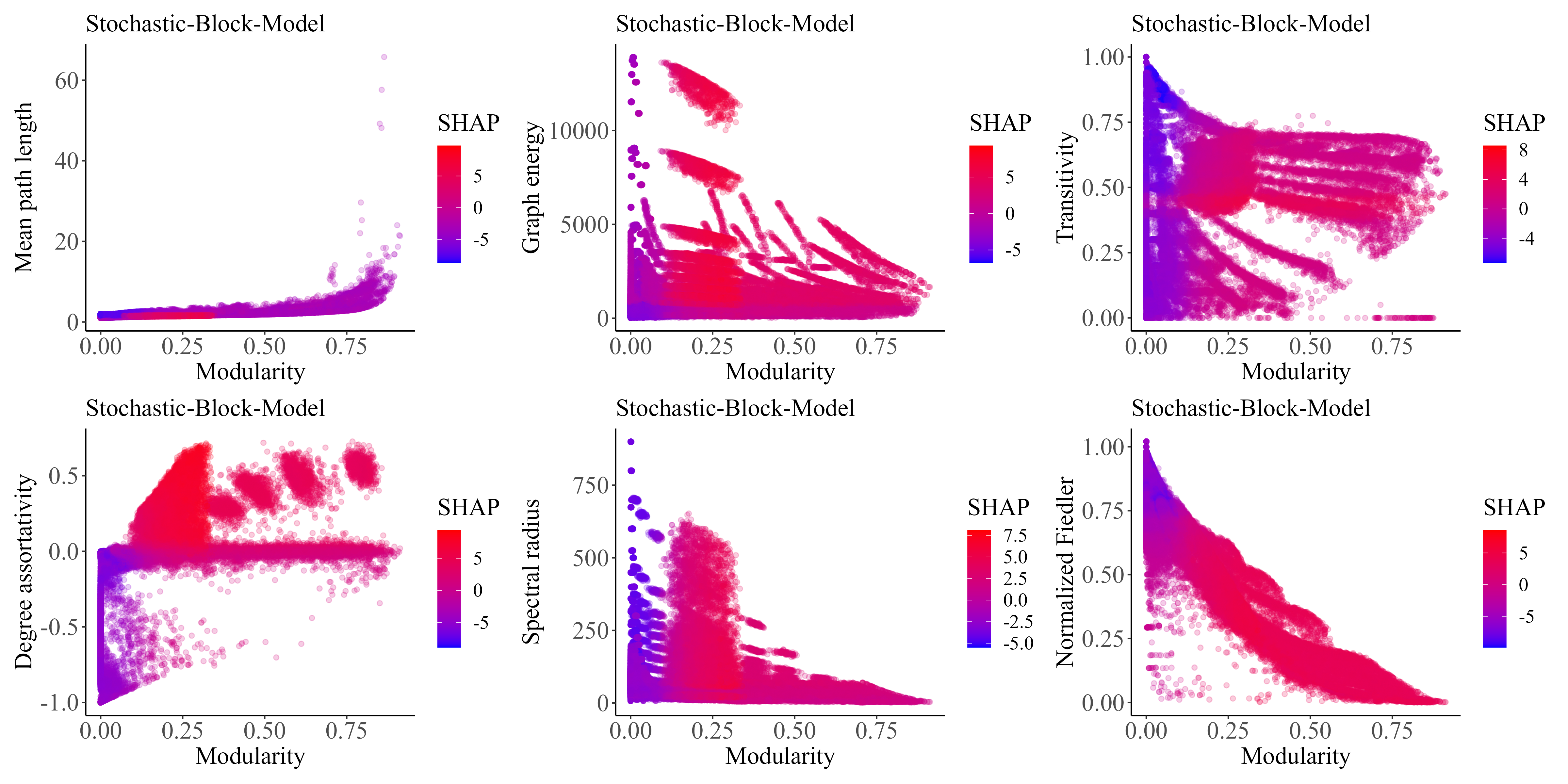}\\
\end{tabular}
 \end{figure*}
 \end{landscape}

\begin{landscape}
\begin{figure*}[htbp]
\centering
\begin{tabular}{l}
\includegraphics[scale=0.4]{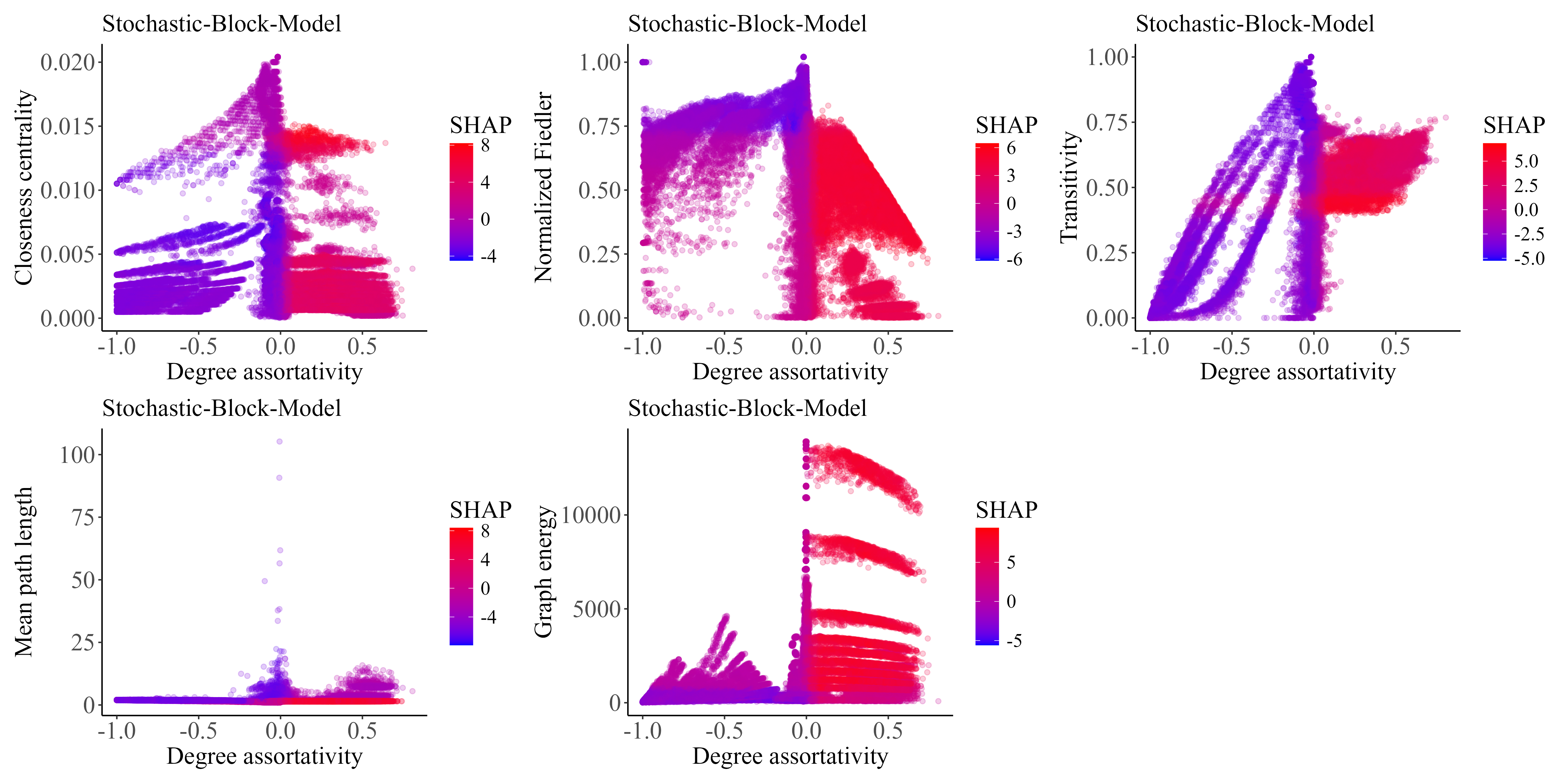}\\[1em]
\end{tabular}
\end{figure*}
\end{landscape}

\begin{landscape}
\begin{figure*}[htbp]
\centering
    \begin{tabular}{l}
     (b) \\
        \includegraphics[scale=0.4]{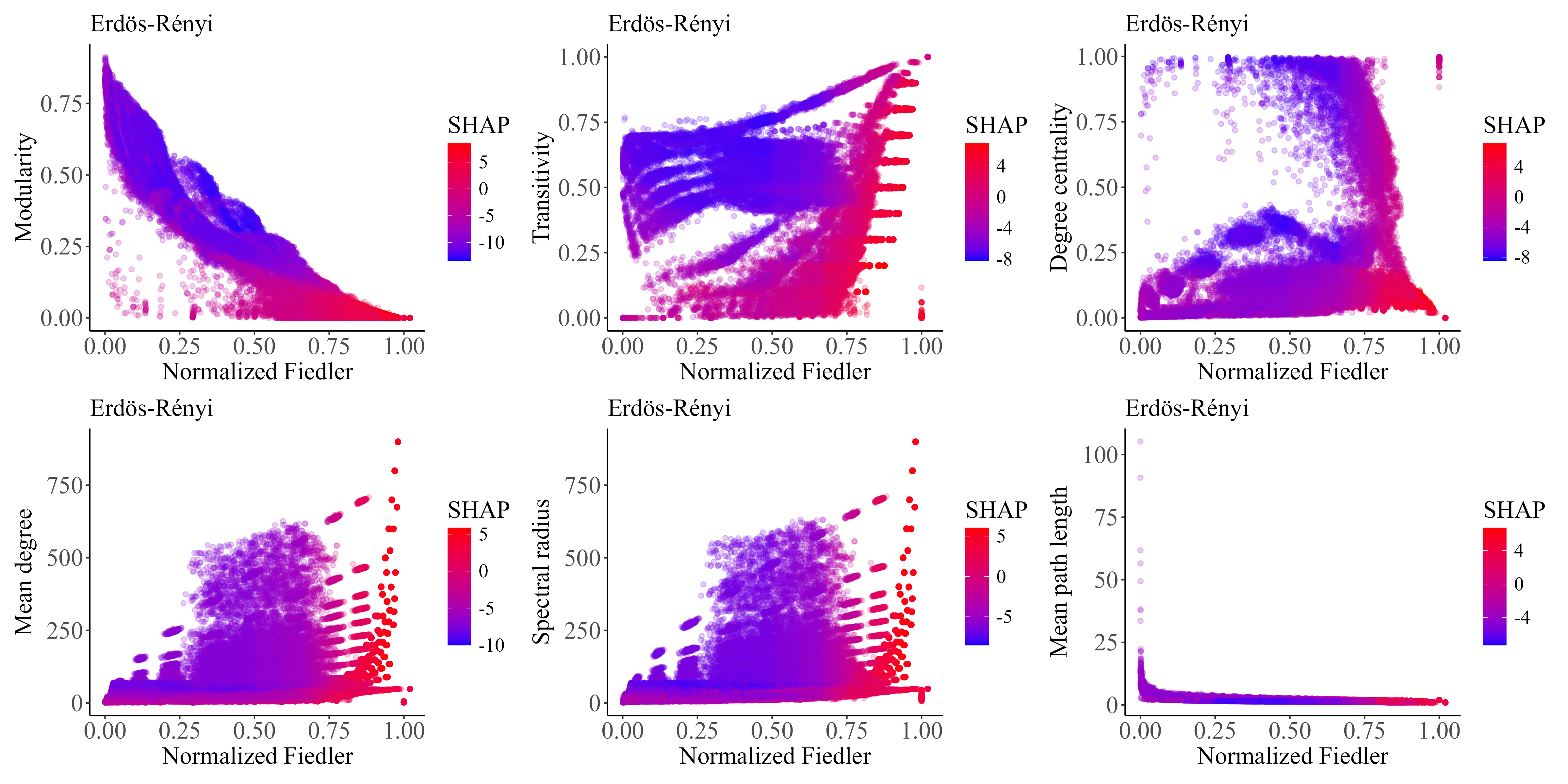} \\[1em]
   \end{tabular}
\end{figure*}
\end{landscape}

\begin{landscape}
\begin{figure*}[htbp]
\centering
    \begin{tabular}{l}
     (c) \\ \includegraphics[scale=0.4]{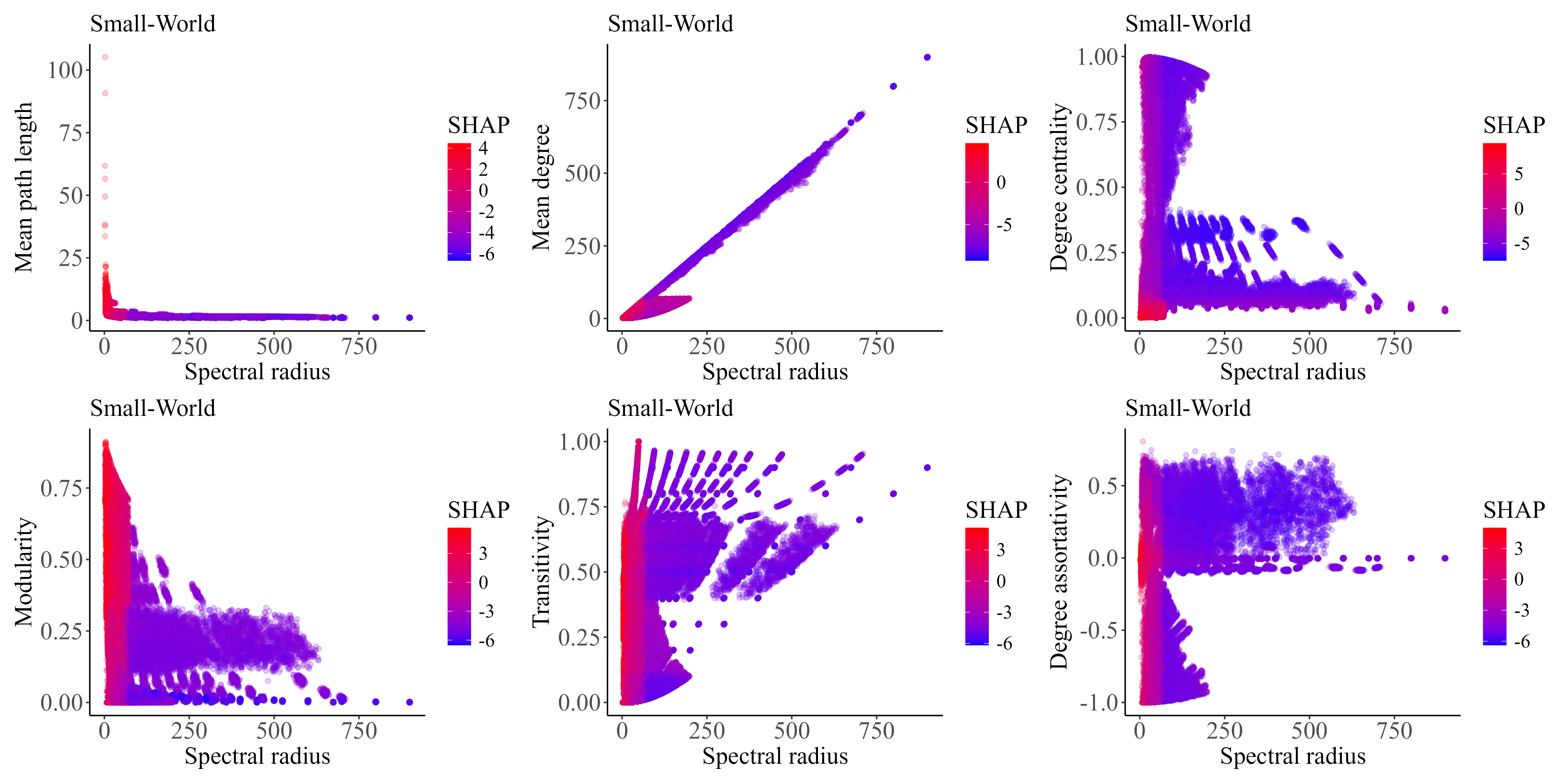} \\[1em]
   \end{tabular}
\end{figure*}
\end{landscape}

\begin{landscape}
\begin{figure*}[htbp]
\centering
\begin{tabular}{l}
    (d) \\
\includegraphics[scale=0.4]{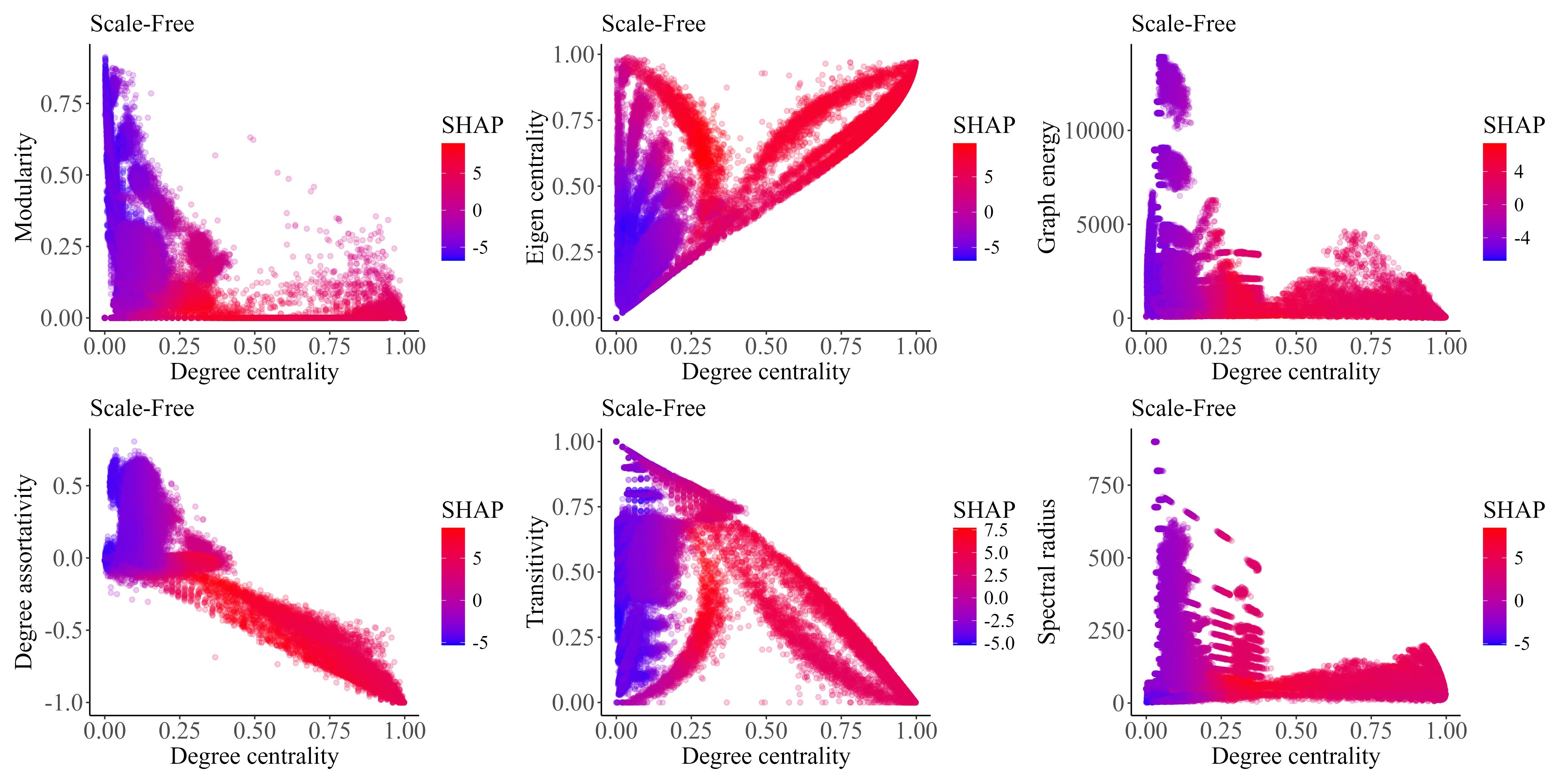}\\[1em]
  \end{tabular}
\end{figure*}
\end{landscape}

\begin{figure*}[htbp]
\centering
	\caption{Two-way interaction effect between pairs of selected network features for (a) stochastic-block-model, (b) Erdös-Rényi, (c) small-world, and (d) scale-free classes.
 These plots shows the combined effect of four selected features (different for each model class) against the most important features for the generative models in their predictions.
 These plots visualize how the predicted outcome of the a) stochastic-block-model, (b) Erdös-Rényi, (c) small-world, and (d) scale-free classes depends on the values of two input features simultaneously.
 The plots represents the prediction of the generative models arising from the dependency between the main features (on the x-axis) alongside the different four selected features (on the y-axis), which also varies depending on the generative model.
 The color or shading of the plots (manually set) indicates the predicted outcome or class probability for the generative model types.
 Overall, this plot shows at which features values the combined effect of two features existing simultaneously positively, or adversely affect the generative models class prediction.
 The gaps in these plots signify the absence of simulated networks for specific combinations of feature values, owing to the stochastic nature of the network synthesis and the use of an unknown parameter space beyond our control}
 \label{fig:sf_er_sw_2d_plot}
\end{figure*}

\begin{figure*}[b]
\centering
\includegraphics[scale=0.3]{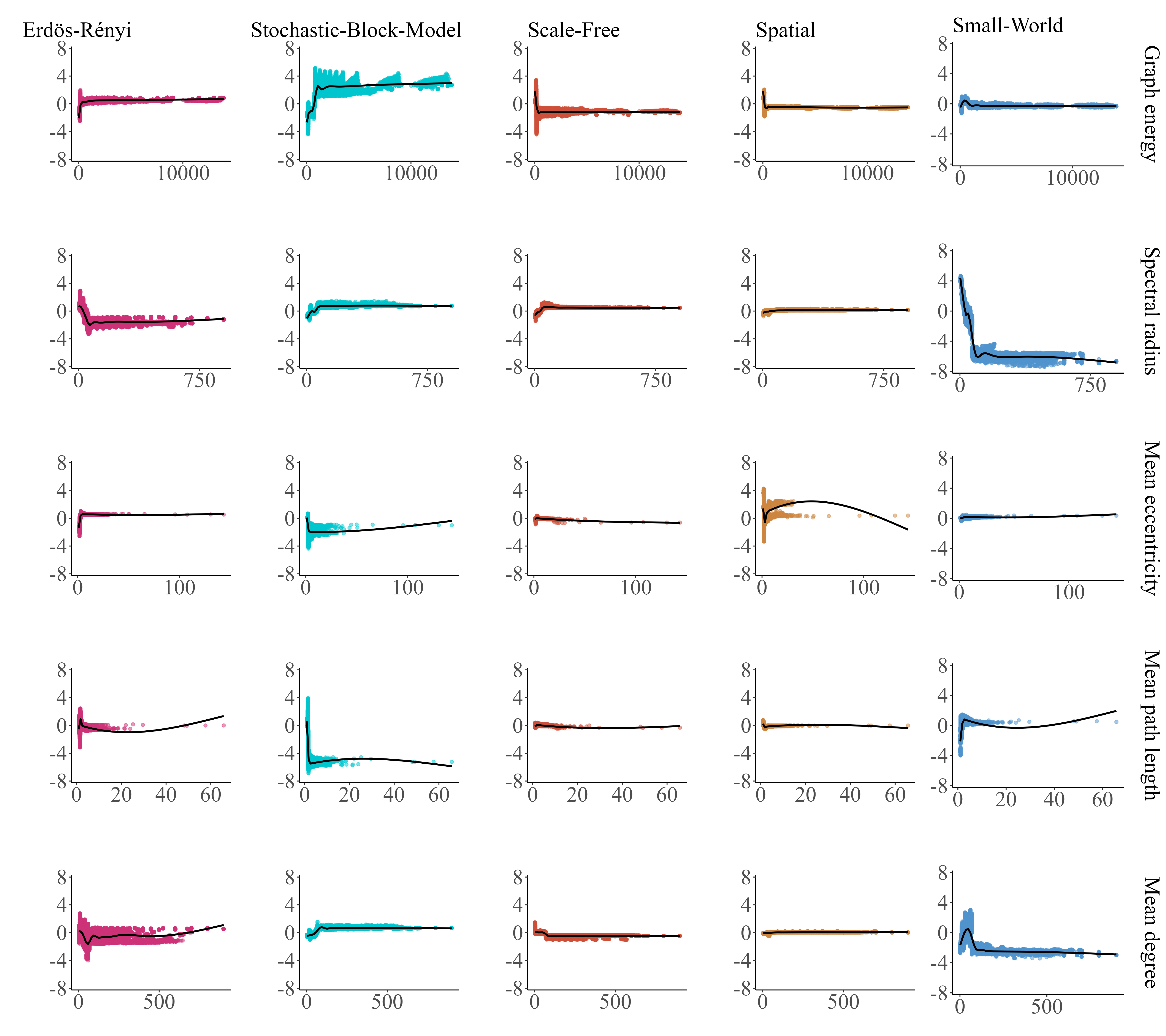}
\end{figure*}

\begin{figure*}[b]
\centering
	\caption{
 SHAP dependency plot showing the scatter plot of the relationship between 
 between the less important features in predicting Erdös-Rényi, stochastic-block-model, scale-free, spatial, and small-world.
 We can see in this plot, each feature and its importance in predicting the final output of the generative models across different values. 
 The y-axis typically represents the SHAP values, which quantifies the impact of the feature on the model's prediction, while the x-axis shows the feature's value.
 This plot provides insights into how the model's prediction changes as the feature's value varies, for each class separately.
 The direction and magnitude of the SHAP values across the different classes aid in discerning how important the features are for each class prediction and whether its effect is consistent across all classes or varies.}
 \label{fig:dependency_plot2}
\end{figure*}

\begin{landscape}
\begin{figure*}[t!]	\centering
\includegraphics[scale=0.3]{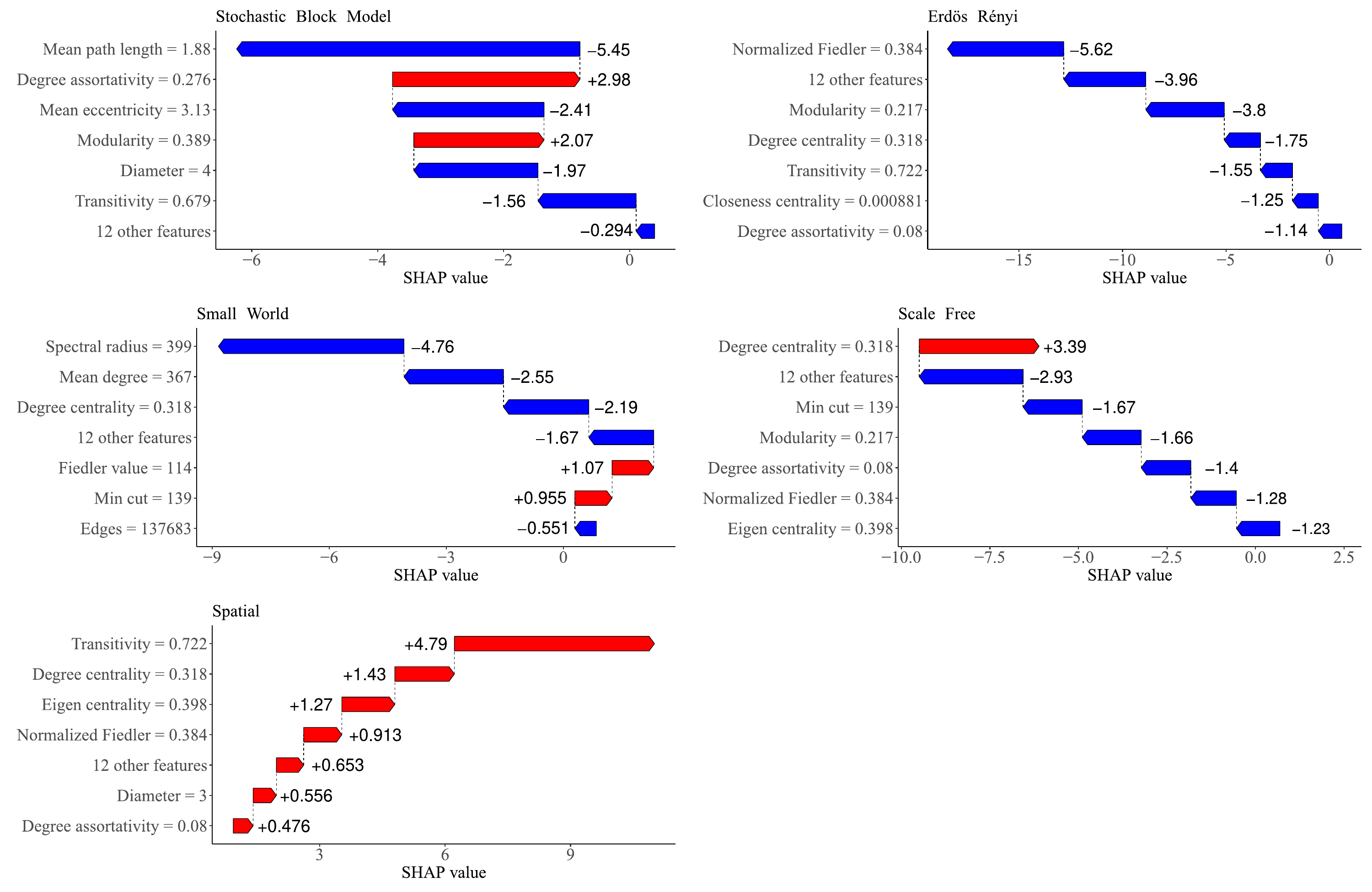}
	\caption{SHAP waterfall plot explaining individual predictions on the five model class categories. This plot shows how each features impact the final model class outputs. Features gradually pushing the model prediction higher (towards the right) are shown in red and those pushing the model prediction lower (towards the left) are shown in blue}
 \label{fig:waterfall}
\end{figure*}
\end{landscape}

\begin{landscape}
\begin{figure*}[htbp]
\centering	\includegraphics[scale=0.5]{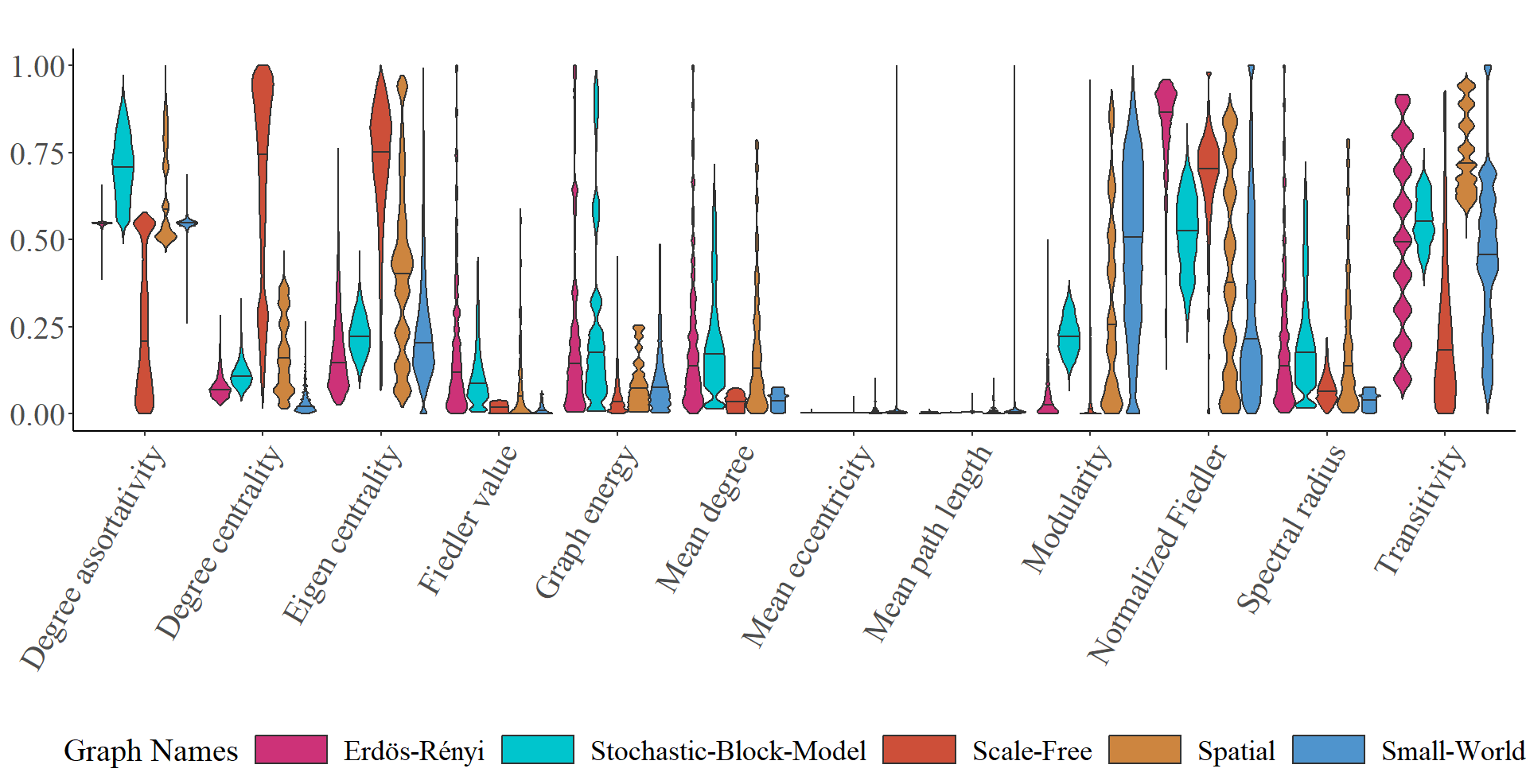}
\caption{Violin plot showing the distribution of the network features across the five model classes.}
 \label{fig:Violin}
\end{figure*}
\end{landscape}

\begin{landscape}
\centering
\begin{table}[h!]
\caption{\bf Range of parameter values used for simulation of data.}  
\label{tab:s1}
\begin{tabular}{llc}
\hline
Models&Structure&Parameters\\
\hline
Erdös-Rényi&random connection& 0.1 $\leq$ $P^{(ER)}$ $\leq$ 0.9\\
Small-world&rewiring connections to random long-range nodes &$P^{(SW)}$ = 0.1, 0.3; 1 $\leq$ $l^{s}$ $\leq$ 35 \\
Scale-free&high-degree nodes& $\alpha$ = 1, 2, 3; 1 $\leq$ m $\leq$ 35\\
Spatial&spatial allocation of nodes&0.1 $\leq$ r $\leq$ 0.9\\
Stochastic-block-model&community embeddings of nodes& 1 $\leq$ $P$ $\leq$ 250\\
\end{tabular}
\end{table}
\end{landscape}
\end{document}